\documentclass[floatfix, preprint, showpacs, showkeys, preprintnumbers, nofootinbib, superscriptaddress]{revtex4-1}
\usepackage[utf8]{inputenc}
\usepackage[sort&compress]{natbib}
\usepackage{ulem}
\usepackage{bm}
\usepackage{times}
\usepackage{amssymb,amsbsy,amsmath,amsfonts}
\usepackage{graphicx}
\usepackage{epstopdf}
\usepackage{float}
\usepackage{color}
\usepackage{morefloats}
\usepackage{rotating}
\usepackage{srcltx}
\usepackage{slashed}
\usepackage{subfigure}
\usepackage{multirow}
\usepackage{verbatim}
\usepackage{hyperref}
\usepackage{overpic}
\usepackage{tabularx}

\begin{document}

\title{Is $X(7200)$ the heavy anti-quark diquark symmetry partner of $ X(3872)$?}

\author{Ming-Zhu Liu}
\affiliation{School of Space and Environment, Beihang University, Beijing 100191, China}
\affiliation{School of Physics, Beihang University, Beijing 100191, China}

\author{Li-Sheng Geng}\email{lisheng.geng@buaa.edu.cn}
\affiliation{School of Physics, Beihang University, Beijing 100191, China}
\affiliation{
Beijing Key Laboratory of Advanced Nuclear Materials and Physics,
Beihang University, Beijing 100191, China}

\affiliation{Beijing Advanced Innovation Center for Big Data-Based Precision Medicine, School of Medicine and Engineering, Beihang University, Beijing, 100191}

\affiliation{School of Physics and Microelectronics, Zhengzhou University, Zhengzhou, Henan 450001, China}
\date{\today}
\begin{abstract}
The $D^{(\ast)}\Xi_{cc}^{(\ast)}$ system  and $\bar{\Xi}_{cc}^{(\ast)}\Xi_{cc}^{(\ast)}$ system can be related to the $D^{(\ast)}\bar{D}^{(\ast)}$ system via heavy anti-quark di-quark symmetry (HADS).  In this work,  we employ a contact-range effective field theory to systematically investigate the likely existence of molecules in these  systems in terms of  the  hypothesis that X(3872)  is a $1^{++}$~$D\bar{D}^{\ast}$ bound state in the isospin symmetry limit, with some of the unknown low energy constants estimated using  the light-meson saturation approximation. In the meson-meson system, a $J^{PC}=2^{++}$~$\bar{D}^{\ast}D^{\ast}$ molecule commonly referred to as $X(4013)$ is reproduced, which is the heavy quark spin partner of $X(3872)$.    In the meson-baryon system, we predict two triply charmed pentaquark molecules, $J^{P}=1/2^{-}$~$D^{\ast}\Xi_{cc}$ and $J^{P}=5/2^{-}$~$D^{\ast}\Xi_{cc}^{\ast}$. In the baryon-baryon system, there exist seven di-baryon molecules, $J^{PC}=0^{-+}$~$\bar{\Xi}_{cc}\Xi_{cc}$, $J^{PC}=1^{--}$~$\bar{\Xi}_{cc}\Xi_{cc}$, $J^{PC}=1^{-+}$~$\bar{\Xi}_{cc}\Xi_{cc}^{\ast}$, $J^{PC}=1^{--}$~$\bar{\Xi}_{cc}\Xi_{cc}^{\ast}$,
$J^{PC}=2^{-+}$~$\bar{\Xi}_{cc}\Xi_{cc}^{\ast}$, $J^{PC}=2^{-+}$~$\bar{\Xi}_{cc}^{\ast}\Xi_{cc}^{\ast}$
and $J^{PC}=3^{--}$~$\bar{\Xi}_{cc}^{\ast}\Xi_{cc}^{\ast}$.  Among them, the $J^{PC}=0^{-+}$~$\bar{\Xi}_{cc}\Xi_{cc}$ and/or $J^{PC}=1^{--}$~$\bar{\Xi}_{cc}\Xi_{cc}$ molecules may contribute to the $X(7200)$ state recently observed  by the LHCb Collaboration,  which implies that $X(7200)$ can be related to $X(3872)$ via HADS.  As a byproduct, with the heavy quark flavor symmetry we also study likely existence of molecular states in the $B^{(\ast)}\bar{B}^{(\ast)}$, $\bar{B}^{(\ast)}\Xi_{bb}^{(\ast)}$, and $\bar{\Xi}_{bb}^{(\ast)}\Xi_{bb}^{(\ast)}$ systems.
\end{abstract}


\maketitle

\section{Introduction}

Hadrons  can be classified either as mesons made of a pair of quark and antiquark or as baryons made of three quarks in the conventional quark model,  which has been rather successful at least up to 2003.
According to the  quark model~\cite{GellMann:1964nj}, more complicated quark configurations, such as multiquark states, should also exist, which are also allowed  by QCD.
In recent years a lot of $XYZ$ and $P_{c}$ states that can not easily  fit into the conventional quark model were discovered by many collaborations all over the world,  which not only  challenges the conventional quark model  but also opens a  new era for hadron physics~\cite{Chen:2016qju,Hosaka:2016ypm,Guo:2017jvc,Olsen:2017bmm,Brambilla:2019esw}.

In 2003, the Belle Collaboration reported a resonant state, $X(3872)$,  in the  $J/\psi \pi^{+}\pi^{-}$ invariant mass spectrum of the  $\Lambda_{b}\rightarrow J/\psi \pi^{+}\pi^{-} K$ decay~\cite{Aaij:2015tga}, which has later been confirmed by other collaborations~\cite{Aubert:2005rm,Acosta:2003zx,Abazov:2004kp,Aubert:2004ns,Ablikim:2013dyn,Aaij:2011sn}.
Its spin-parity was determined by the LHCb Collaboration in 2013~\cite{Aaij:2013zoa}.
 Being close to the mass threshold of $D\bar{D}^{\ast}$, it is naturally considered to be a hadronic molecule.
 In addition, another convincing evidence of the molecular nature of $X(3872)$ is its isospin breaking decays into $J/\Psi \pi\pi$ and $J/\Psi \pi\pi\pi$~\cite{Abe:2005ix,delAmoSanchez:2010jr}.
 A series of theoretical works  interpreted $X(3872)$ as a $J^{PC}=1^{++}$~$\bar{D}D^{\ast}$ molecule~\cite{Swanson:2003tb,Voloshin:2003nt,AlFiky:2005jd,Liu:2008fh,Sun:2011uh,Nieves:2012tt,Guo:2013sya,Karliner:2015ina,Liu:2019stu}. Although  the molecular interpretation seems to be the most popular, its nature has not been completely unveiled.~\footnote{Other explanations also exist such as compact tetraquark~\cite{Maiani:2004vq,Ebert:2005nc,Matheus:2006xi,Wang:2013vex}, charmonium~\cite{Barnes:2003vb,Vijande:2004he}, molecule-charmonium~\cite{Kalashnikova:2005ui,Matheus:2009vq,Ferretti:2013faa}, and so on.  }
 Nonetheless, given current experimental and theoretical results,  one may conclude that the molecule components of $X(3872)$ must play a relevant (if not dominant) role.
 In our previous work~\cite{Liu:2019stu}, assuming that $X(3872)$ is a $J^{PC}=1^{++}$~$\bar{D}D^{\ast}$ molecule,  we have used the one boson exchange model to predict  the binding energies and scattering lengths of the heavy quark spin symmetry (HQSS) partners of $X(3872)$.

HQSS  plays an important role in describing the spectra of heavy hadrons including both conventional and exotic ones. The $D$ and $D^{\ast}$ mesons as well as $\Sigma_{c}$ and $\Sigma_{c}^{\ast}$ baryons can be regarded as doublets of HQSS. Applying HQSS to their bottom partners, the mass splittings of doublets become small, which indicates that HQSS works well in the heavy quark limit. The interactions between heavy hadrons   are also constrained by HQSS, which  can further decrease the number of unknown couplings  and increase the predictive power of effective field theories (EFT) as well. In the meson-meson system, $D_{s0}^*(2317)$ and $D_{s1}(2460)$ are assigned as $DK$ and $D^{\ast}K$ hadronic molecules, respectively, belonging to a HQSS doublet ~\cite{Barnes:2003dj,Kolomeitsev:2003ac,Guo:2006fu,Gamermann:2006nm,Faessler:2007gv,Faessler:2007us,Liu:2012zya,Altenbuchinger:2013vwa}.
 A lot of theoretical works described $Z_{c}(3900)$ and $Z_{c}(4020)$  as  isovector resonances generated by the  $\bar{D}D^{\ast}$ and $\bar{D}^{\ast}D^{\ast}$ interactions, respectively, which indicates that  $Z_{c}(3900)$ can be related to $Z_{c}(4020)$ via HQSS~\cite{Guo:2013sya,Dong:2013iqa,Wang:2020dko}. Very recently  $Z_{cs}(3985)$ was proposed to be the $SU(3)$-flavor partner of $Z_{c}(3900)$~\cite{Yang:2020nrt,Meng:2020ihj,Du:2020vwb}.  In the heavy meson and baryon systems, $P_{c}(4312)$, $P_{c}(4440)$, and $P_{c}(4457)$ are nicely arranged into a multiplet of  $\bar{D}^{(\ast)}\Sigma_{c}^{(\ast)}$  hadronic molecules~\cite{Liu:2019tjn,Du:2019pij,Xiao:2019aya,Sakai:2019qph,Yamaguchi:2019seo,Lin:2019qiv,Valderrama:2019chc,Liu:2019zvb}, and in terms of HQSS  there should  exist other four states which need to be confirmed by future experiments.

Heavy quark flavor symmetry has already shown its power in constructing the spectra of charmed and bottom hadrons. In short, if there exists a charmed hadron there should also exist the corresponding   bottom partner.  Heavy anti-quark di-quark symmetry(HADS) dictates that in terms of  color degree of freedom a pair of heavy quarks can be related to a heavy antiquark in the heavy quark limit~\cite{Savage:1990di}. From this symmetry we can derive the following mass relation: $m_{\Xi_{cc}^{\ast}}-m_{\Xi_{cc}}=\frac{3}{4}(m_{D^{\ast}}-m_{D})$~\cite{Hu:2005gf}. Although this relation  has not been confirmed by experiments so far,  it is consistent with lattice QCD calculations and phenomenological studies~\cite{Padmanath:2015jea,Mathur:2018rwu,Roncaglia:1995az,Karliner:2014gca}. In our previous work, we have employed HADS to extend the $\bar{D}^{(\ast)}\Sigma_{c}^{(\ast)}$ system to the $\Xi_{cc}^{(\ast)}\Sigma_{c}^{(\ast)}$ system, and predicted the likely existence of $\Xi_{cc}^{(\ast)}\Sigma_{c}^{(\ast)}$ bound states according to the hypothesis that $P_{c}(4440)$ and $P_{c}(4457)$ are $\bar{D}^{\ast}\Sigma_{c}$ bound states~\cite{Pan:2019skd,Pan:2020xek}.

\begin{figure}[!h]
 \center{\includegraphics[width=14cm]  {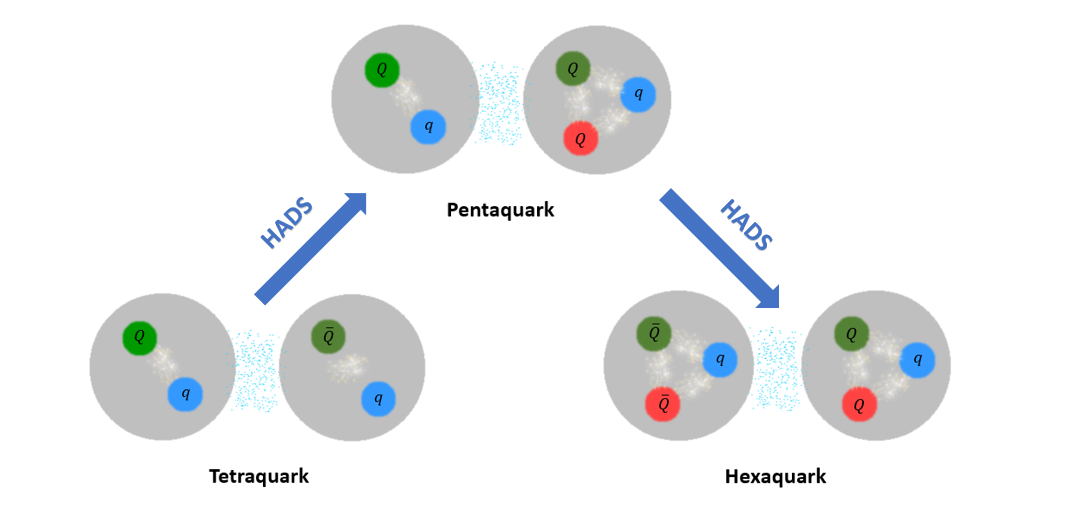}}
 \caption{\label{cha}HADS related hidden charm tetraquark states, triply charmed pentaquark states, and hidden charmed hexaquark states.}
 \end{figure}

Along the line of Ref.~\cite{Pan:2019skd}, we can extend the $D^{(\ast)}\bar{D}^{(\ast)}$ system to the $D^{(\ast)}\Xi_{cc}^{(\ast)}$ system and then to the $\bar{\Xi}_{cc}^{(\ast)}\Xi_{cc}^{(\ast)}$ system via HADS as shown in Fig. \ref{cha}, and systematically investigate the  likely existence of bound states in these systems using a contact-range  EFT. In Refs.~\cite{Guo:2013sya,Guo:2013xga} the authors have already studied the $D^{(\ast)}\bar{D}^{(\ast)}$ and  $D^{(\ast)}\Xi_{cc}^{(\ast)}$ systems in the contact-range EFT.  However, because of the existence of two unknown couplings and only one effective input,  the full spectra of these two systems could not be determined. In this work we employ the light-meson saturation approach to estimate the ratio of the two couplings~\cite{Peng:2020xrf}.
In our previous work~\cite{Liu:2019zvb}, we found that the two couplings of the $\bar{D}^{(\ast)}\Sigma_{c}^{(\ast)}$ system determined by the contact-range EFT and the light-meson saturation approach are quite similar.
 In Ref.~\cite{Peng:2020hql} Peng. et al used the light-meson saturation approach to estimate the contact-range potential between $\bar{D}\Xi_{c}$ and $\bar{D}\Xi_{c}^{\prime}$,  yielding results consistent with the local hidden gauge approach~\cite{Xiao:2019gjd,Liu:2020hcv}. Therefore, in this work, to fully determine the mass spectra of these systems, we will combine  the hypothesis that $X(3872)$ is a $J^{PC}=1^{++}$ bound state and the light-meson saturation approach  to determine the values of the only two unknown couplings $C_{a}$ and $C_{b}$, then calculate the binding energies (if bound states exist) and scattering lengths of the $D^{(\ast)}\bar{D}^{(\ast)}$,  $D^{(\ast)}\Xi_{cc}^{(\ast)}$,  and $\bar{\Xi}_{cc}^{(\ast)}\Xi_{cc}^{(\ast)}$ systems.

Recently, the LHCb Collaboration reported one resonant state with a mass of 6.9 GeV in the $J/\psi $-$J/\psi$ invariant mass spectrum, labelled as $X(6900)$. In addition, there exist two more structures, a broad one from 6.2 to 6.8 GeV and a vague structure around 7.2 GeV~\cite{Aaij:2020fnh}. These tetraquark states  containing fully charmed quarks  undoubtedly open another new era for hadron physics since the discovery of $X(3872)$, which has motivated a lot of theoretical studies~\cite{Ma:2020kwb,Dong:2020nwy,Wang:2020dlo,Karliner:2020dta,Maciula:2020wri,Chao:2020dml,Richard:2020hdw,Maiani:2020pur,Sonnenschein:2020nwn,Giron:2020wpx,Wang:2020gmd,Weng:2020jao,Zhu:2020xni,Guo:2020pvt,Zhu:2020snb,Cao:2020gul,liu:2020eha,Gong:2020bmg}.
Two popular interpretation of these structures are resonances generated by several coupled channels of charmonium pairs and compact tetraquark states of diquark and anti-diquark. In this work we mainly focus on the structure near 7.2 GeV, namely $X(7200)$, whose mass is close to the mass threshold of $\bar{\Xi}_{cc}\Xi_{cc}$.    Since the $\bar{\Xi}_{cc}^{(\ast)}\Xi_{cc}^{(\ast)}$ system can be related to the $\bar{D}^{(\ast)}D^{(\ast)}$ system via HADS, we further explore whether  there exist $\bar{\Xi}_{cc}\Xi_{cc}$ molecules from the hypothesis that $X(3872)$ is a $J^{PC}=1^{++}$ bound state, in other words, whether $X(7200)$ can be related to $X(3872)$ via HADS.

  The manuscript is structured as follows. In Sec.~\ref{sec:obe} we present  the contact-range potentials of the $D^{(\ast)}\bar{D}^{(\ast)}$,  $D^{(\ast)}\Xi_{cc}^{(\ast)}$,  and $\bar{\Xi}_{cc}^{(\ast)}\Xi_{cc}^{(\ast)}$ systems.
In Sec.~\ref{sec:pre}
we  introduce the light-meson saturation approach to help determine the unknown couplings. Then we calculate the binding energies (if bound states exist) and scattering lengths of the $D^{(\ast)}\bar{D}^{(\ast)}$,  $D^{(\ast)}\Xi_{cc}^{(\ast)}$  and $\bar{\Xi}_{cc}^{(\ast)}\Xi_{cc}^{(\ast)}$ systems.
Finally we present the conclusions in Sec.~\ref{sum}

\section{Formalism}
\label{sec:obe}
Generically, the interaction between two heavy hadrons can be decomposed into a long- and a short-range piece within the EFT. The long-range piece corresponds to the one-pion-exchange potential, while the short-range piece can be described by a series of contact-range potentials with unknown couplings.  In the present context, the short-range interaction between two heavy hadrons is constrained by heavy quark symmetry. In the following we explain how to derive the   $D^{(\ast)}\bar{D}^{(\ast)}$,  $D^{(\ast)}\Xi_{cc}^{(\ast)}$,  and $\bar{\Xi}_{cc}^{(\ast)}\Xi_{cc}^{(\ast)}$ interactions. In the line of Refs.~ \cite{Liu:2020nil,Pan:2019skd}, their interactions can be determined in the EFT approach. One should note that we just consider the leading order contact-range potentials because our previous studies indicated that the  pion exchanges
are perturbative in the charm sector~\cite{Valderrama:2012jv,Lu:2017dvm}.
The short-range interaction of the meson-meson system can be written as
\begin{eqnarray}
\mathcal{L}_{HH}=C_{a}Tr[H^{\dag}H]Tr[H^{\prime \dag} H^{\prime}]+ C_{b}Tr[H^{\dag}\sigma_{i}H]Tr[H^{\prime \dag}\sigma_{i} H^{\prime}],
\end{eqnarray}
where $C_{a}$ and $C_{b}$ are two unknown couplings and
$H$ is a non-relativistic superfield with $H=\frac{1}{\sqrt{2}}[P+\vec{P}^{\ast}\cdot\vec{\sigma}]$, and
$P$ and $P^{\ast}$ denote pseuoscalar and vector charmed meson fields.
The $D^{(\ast)}\Xi_{cc}^{(\ast)}$ system can be related to the $D^{(\ast)}\bar{D}^{(\ast)}$ system through  HADS, resulting in the following Lagrangian
\begin{eqnarray}
\mathcal{L}_{HT}=C_{a}Tr[H^{\dag}H]Tr[T^{\prime \dag} T^{\prime}]+ C_{b}Tr[H^{\dag}\sigma_{i}H]Tr[T^{\prime \dag}\sigma_{i} T^{\prime}],
\end{eqnarray}
where $T=\frac{1}{\sqrt{3}}\vec{\sigma}\Xi_{cc}+\vec{\Xi}_{cc}^{\ast}$ is a non-relativistic superfield of $\Xi_{cc}$ and $\Xi_{cc}^{\ast}$ respecting HQSS. The Lagrangian of the  $\bar{\Xi}_{cc}^{(\ast)}\Xi_{cc}^{(\ast)}$ system have the same form as those of the  $D^{(\ast)}\bar{D}^{(\ast)}$ and  $D^{(\ast)}\Xi_{cc}^{(\ast)}$ systems
\begin{eqnarray}
\mathcal{L}_{TT}=C_{a}Tr[T^{\dag}T]Tr[T^{\prime \dag} T^{\prime}]+ C_{b}Tr[T^{\dag}\sigma_{i}T]Tr[T^{\prime \dag}\sigma_{i} T^{\prime}].
\end{eqnarray}
The term $C_{a}$ is independent of the spin, while $C_{b}$ accounts for the spin spin interaction.
Using the same approach of Ref.~\cite{Pan:2019skd}, the contact potentials of the $D^{(\ast)}\bar{D}^{(\ast)}$,  $D^{(\ast)}\Xi_{cc}^{(\ast)}$,  and $\bar{\Xi}_{cc}^{(\ast)}\Xi_{cc}^{(\ast)}$ systems  can be easily constructed and are shown in  Table~\ref{tab:potential}.
\begin{table}[!h]
  \centering \caption{Contact-range potentials
    for the heavy meson-heavy anti-meson, heavy meson-doubly heavy baryon, and doubly heavy
    baryon-doubly heavy anti-baryon systems depending on two unknown coupling constants, $C_{a}$ and $C_{b}$.
    These coupling constants can be determined from the sum of $C_{a}$ and $C_{b}$ by reproducing the mass of
    $X(3872)$ and the ratio of $C_{a}$ and $C_{b}$ by the light-meson saturation approach. \label{tab:potential}}
\begin{tabular}{ccc|ccc|ccccc}
  \hline\hline state& $J^{PC}$ &V   & state   &$J^{P}$   &V & state   &$J^{PC}$   &V \\

  \hline\multirow{2}{0.8cm} {$D\bar{D}$} &
 \multirow{2}{0.8cm}{$0^{++}$} &\multirow{2}{0.8cm}{$C_{a}$}
  & \multirow{2}{0.8cm}{$D\Xi_{cc}$}
  & \multirow{2}{0.8cm}{$\frac{1}{2}^{-}$}  & \multirow{2}{0.8cm}{$C_{a}$}  &  $\bar{\Xi}_{cc}\Xi_{cc}$  & $0^{-+}$ & $C_{a}-\frac{1}{3}C_{b}$       \\
 &
 &
  &
  &  &   &  $\bar{\Xi}_{cc}\Xi_{cc}$  & $1^{--}$ & $C_{a}+\frac{1}{9}C_{b}$      \\  \hline
 \multirow{4}{2.0cm}{$D^{\ast}\bar{D}$/$D\bar{D}^{\ast}$}&
  &
  &
  &  &   & \multirow{4}{2.4cm}{$\bar{\Xi}_{cc}^{\ast}\Xi_{cc}/\bar{\Xi}_{cc}\Xi_{cc}^{\ast}$}  & $1^{-+}$ & $C_{a}+C_{b}$        \\
  &
 {$1^{++}$} &{$C_{a}+C_{b}$}
  &\multirow{2}{0.8cm}{$D^{\ast}\Xi_{cc}$}
  & $\frac{1}{2}^{-}$  & $C_{a}+\frac{2}{3}C_{b}$  &  & $1^{--}$ & $C_{a}+\frac{1}{9}C_{b}$          \\
  & {$1^{+-}$} &{$C_{a}-C_{b}$} & & $\frac{3}{2}^{-}$    & $C_{a}-\frac{1}{3}C_{b}$   &    & $2^{-+}$ & $C_{a}+C_{b}$          \\

   &
 &
  &
  &  &   &   & $2^{--}$ & $C_{a}-\frac{5}{3}C_{b}$      \\
  \hline

   &
 &
  &
 $D\Xi_{cc}^{\ast}$ & $\frac{3}{2}^{-}$ & $C_{a}$  & \multirow{4}{1.2cm}{$\bar{\Xi}_{cc}^{\ast}\Xi_{cc}^{\ast}$}   & $0^{-+}$ & $C_{a}-\frac{5}{3}C_{b}$      \\
\multirow{3}{0.8cm}{$D^{\ast}\bar{D}^{\ast}$} & $0^{++}$
&$C_{a}-2C_{b}$  &
\multirow{3}{0.8cm}{$D^{\ast}\Xi_{cc}^{\ast}$}  &$\frac{1}{2}^{-}$
& $C_{a}-\frac{5}{3}C_{b}$  &    &  $1^{--}$  &  $C_{a}-\frac{11}{9}C_{b}$
\\
& {$1^{+-}$}
& {$C_{a}-C_{b}$} &
& $\frac{3}{2}^{-}$ & $C_{a}-\frac{2}{3}C_{b}$  &    &  $2^{-+}$  &  $C_{a}-\frac{1}{3}C_{b}$
\\  & $2^{++}$ & {$C_{a}+C_{b}$} &
& $\frac{5}{2}^{-}$ & $C_{a}+C_{b}$  &    &  $3^{--}$  &  $C_{a}+C_{b}$
\\
\hline\hline
\end{tabular}
\end{table}

To make concrete predictions for likely existence of molecular states we have
to solve a non-relativistic bound state equation with the
contact-range potentials of Table \ref{tab:potential}.
In the momentum space,  we can solve the Lippmann-Schinwinger equation to find  bound state as
\begin{eqnarray}
\phi(k)+\int \frac{d^{3}p}{(2\pi)^3}\langle
k|V|p\rangle\frac{\phi(p)}{B+\frac{p^2}{2\mu}}=0, \label{10}
\end{eqnarray}
where $\phi(k)$ is the vertex function, $B$ the binding energy, and $\mu$
the reduced mass. To solve the equation we  regularize the
contact potential in the following way
\begin{eqnarray}
\langle p| V_{\Lambda}|
p^{\prime}\rangle=C(\Lambda)f(\frac{p}{\Lambda})f(\frac{p^{\prime}}{\Lambda}),
\end{eqnarray}
with $\Lambda$ the cutoff, $f(x)$ a regular function, and $C({\Lambda})$ the
running coupling constant.  A typical choice of the cutoff is
$\Lambda=0.5-1$ GeV, while for the regulator we choose  a Gaussian type
$f(x)=e^{-x^2}$. In this work, to be consistent with the cutoff adopted in the light-meson saturation approach, we use a  cutoff of $\Lambda=0.7$ GeV. Considering only  $S$-wave
contact interactions, the integral equation simplifies
to
\begin{eqnarray}
1+C(\Lambda)\frac{\mu}{\pi^2}\int_{0}^{\infty}dqe^{-2\frac{q^2}{\Lambda^{2}}}\frac{q^2}{B+\frac{\vec{q}^{2}}{2\mu}}=0.
\label{10}
\end{eqnarray}

\section{Numerical results and discussions}
\label{sec:pre}

In this section we study the likely existence of molecular states in the $D^{(\ast)}\bar{D}^{(\ast)}$, $D^{(\ast)}\Xi_{cc}^{(\ast)}$, and $\Xi_{cc}^{(\ast)}\bar{\Xi}_{cc}^{(\ast)}$ systems. The  contact-range potentiasl of these systems are only dependent on two couplings, $C_{a}$ and $C_{b}$.
To determine these couplings, the best approach is to fit to the experimental data.
However, there exists only one molecular candidate $X(3872)$   to determine the sum of  $C_{a}$ and $C_{b}$.   We, therefore, resort to the light-meson saturation approach to estimate the ratio of $C_{a}$ and $C_{b}$.

The assumption of light-meson saturation implies that the values of the EFT couplings are saturated by light meson exchanges.
 Following the formalism of Ref.~\cite{Peng:2020xrf}, we estimate the couplings of $C_{a}$ and $C_{b}$ by the sigma and vector meson exchange saturation,
\begin{eqnarray}
C_{a}^{sat}(\Lambda\sim m_{\sigma},m_{V})&\propto& C_{a}^{S}+ C_{a}^{V},    \\ \nonumber
C_{b}^{sat}(\Lambda\sim m_{\sigma},m_{V})&\propto& C_{b}^{V},
\end{eqnarray}
where the cutoff $\Lambda\sim m_{\sigma},m_{V}$ implies that the saturation works at an EFT cutoff close to the masses of exchanged mesons, i.e., $0.6\sim 0.8$ GeV. The values of saturated couplings are expected to be proportional to the potentials of light meson exchanges in the OBE model once we have removed the spurious Dirac-delta potential.  Thus $C_{a}$ and $C_{b}$ can be written as
\begin{eqnarray}
C_{a}^{sat(\sigma)}(\Lambda\sim m_{\sigma})&\propto& -\frac{g_{\sigma}^2}{m_{\sigma}^2},    \\ \nonumber
C_{a}^{sat (V)}(\Lambda\sim m_{V})&\propto& -\frac{g_{v}^2}{m_{v}^2}(1+\vec{\tau}_{1}\cdot\vec{\tau}_{2}),  \\ \nonumber
C_{b}^{sat (V)}(\Lambda\sim m_{V})&\propto& -\frac{f_{v}^2}{4 M^2}(1+\vec{\tau}_{1}\cdot\vec{\tau}_{2}),
\label{123}
\end{eqnarray}
where $g_{\sigma_{1}}$ denotes the charm meson coupling to sigma meson, $g_{v1}$ and $f_{v1}$ denote electric-type and magnetic-type couplings between charm mesons and light vector mesons, and $M$ is a mass scale to render $f_{v1}$ dimensionless, see, e.g., Table \ref{tab:couplings2}.
The proportionality constant is unknown and depends on the details of the renormalization procedure. However, assuming that the constant is the same for $C_{a}^{sat}$ and $C_{b}^{sat}$, we can calculate their ratio.

\begin{table}[ttt]
\centering
\caption{Masses of light mesons, heavy mesons, and heavy baryons as well as couplings of the heavy mesons to the light mesons. The magnetic couplings of vector mesons are defined as $f_{v}=\kappa_{v}g_{v}$, and
$M$ refers to the mass scale (in MeV) involved in the magnetic-type couplings.
}
\label{tab:couplings2}
\begin{tabular}{cc|cc|cc|ccccccc}
  \hline \hline
  Coupling  & Value for $D^{(\ast)}$   &  Light meson  & M(MeV)  & Heavy meson  & M(MeV)   &  Heavy baryon  & M(MeV)  \\
  \hline
  $g_{\sigma }$ & 3.4 &    $\pi$ & 138 & $D$ & 1867 & $\Xi_{cc}$  & 3621 \\
  $g_{v }$ & 2.6 &   $\sigma$ & 600  & $D^{\ast}$ &2009  & $\Xi_{cc}^{\ast}$  & 3727\cite{Hu:2005gf}\\
  $\kappa_{v }$ & 2.4 &  $\rho$ & 770 & $B$ & 5279   &  $\Xi_{bb}$ &  10127\cite{Lewis:2008fu} \\
  $M$ & 940 &  $\omega$ & 780  & $B^{\ast}$ & 5325 & $\Xi_{bb}^{\ast}$ &  10151\cite{Lewis:2008fu}\\
  \hline \hline
\end{tabular}
\end{table}

In the following, we explain how to determine the values of $C_{a}$ and $C_{b}$ in detail.
Assuming that $X(3872)$ is a $J^{PC}=1^{++}$~ $D^{\ast}\bar{D}$ bound state with the contact potential $C_{a}+C_{b}$, we can determine the sum of $C_{a}$ and $C_{b}$ as 31.4 GeV$^{-2}$ by  Eq. (\ref{10}), where the cutoff is fixed at 0.7 GeV. With this sum we can only predict the binding energies of states for which the contact potentials are $C_{a}+C_{b}$.  Fortunately, combining the ratio of $C_{a}$ and $C_{b}$ determined by the light-meson saturation, we can determine the values of $C_{a}$ and $C_{b}$. The ratio is
\begin{eqnarray}
\frac{C_{b}^{sat}}{C_{a}^{sat}}=\frac{C_{b}^{sat (V)}}{C_{a}^{sat(V+\sigma)}}\approx 0.35.
\end{eqnarray}
Comparing the ratio $C_{b}/C_{a}\approx 0.134$ in the meson-baryon system of Ref. \cite{Liu:2019zvb},   we find that the spin-spin term plays a more important role in the meson-meson system, which could be the main reason why we can not obtain a complete multiplet of hadronic molecules in the heavy meson-meson system.
Using the ratio determined by  the light-meson saturation we finally obtain $C_{a}=-23.3$ GeV$^{-2}$ and $C_{b}=-8.1$ GeV$^{-2}$, which can help us compute the full spectra of the $D^{(\ast)}\bar{D}^{(\ast)}$, $D^{(\ast)}\Xi_{cc}^{(\ast)}$,  and $\bar{\Xi}_{cc}^{(\ast)}\Xi_{cc}^{(\ast)}$ systems in the EFT approach. According to the EFT, the breaking of HADS is of the order of $\Lambda_{QCD}/(m_{Q}v)$~\cite{Savage:1990di}, where $\Lambda_{QCD}\sim200-300$ MeV and $m_{Q}$ and $v$ are the mass and velocity of the di-quark.
From the estimation of Ref. \cite{Cho:1992cf}, $m_{Q}v\sim$0.8 GeV for a charm quark pair. In this work we take $30\%$ uncertainty for HADS. The breaking of HQSS is taken  to be $15\%$~\cite{Pan:2019skd}.

\subsection{ Meson-meson system}
\begin{table}[!h]
\centering
\caption{ Scattering lengths ($a$ in fm), binding energies ($B$ in MeV if bound states exist) and mass spectra ($M$ in MeV) of prospective isoscalar  heavy antimeson-meson molecules. The $c$ and $b$ subscripts denote the charm and bottom sector, respectively.  The uncertainties originate from the HQSS breaking of the order $15\%$.
  }
\label{tab:comparison-EFT1}
\begin{tabular}{c|cc|ccc|c|ccc}
\hline\hline
molecule & $I$ & $J^{PC}$  & $a_{c}$ (fm)
& $B_{c}$ (MeV)
& $M_{c}$ (MeV )   & molecule  &$a_{b}$ (fm)
& $B_{b}$ (MeV)
& $M_{b}$ (MeV ) \\
  \hline
  $D\bar{D}$ & $0$ & $0^{++}$ &$-20.5_{+27.5}^{+17.2}$ & $\dag$ & $\dag$& $B\bar{B}$  &$1.1_{-0.1}^{+0.1}$ & $20.1^{+7.3}_{-6.7}$ & $10537.9$ \\
  \hline
  $D \bar{D}^{\ast}+ D^{\ast}\bar{D}$ & $0$ & $1^{++}$
  & $2.7_{-0.7}^{+2.8}$  & Input & Input & $B \bar{B}^{\ast}+ B^{\ast}\bar{B}$& $1.0_{+0.0}^{-0.0}$  & $37.9^{+10.9}_{-10.4}$ & 10566.1\\
  $D \bar{D}^{\ast}- D^{\ast}\bar{D}$ & $0$ & $1^{+-}$ & $-1.3_{-0.8}^{+0.4}$  & $\dag$ & $\dag$  & $B \bar{B}^{\ast}- B^{\ast}\bar{B}$& $1.6_{-0.2}^{+0.5}$  & $5.7^{+3.5}_{-2.9}$ & $10598.3$ \\
  \hline
  $D^{\ast}\bar{D}^*$ & $0$ & $0^{++}$  & $-0.3^{-0.1}_{+0.1}$   & $\dag$ & $\dag$ &$B^{\ast}\bar{B}^*$ & $-3.1^{+1.5}_{-6.9}$   & $\dag$ & $\dag$ \\
  $D^{\ast}\bar{D}^*$ & $0$ & $1^{+-}$    &$-1.5_{-1.0}^{+0.5}$  &$\dag$ & $\dag$ &$B^{\ast}\bar{B}^*$  &$1.6^{+0.5}_{-0.2}$  &$5.8_{-3.0}^{+3.5}$ & $10644.2$\\
   $D^{\ast}\bar{D}^*$ & $0$ & $2^{++}$  & $2.5_{-0.6}^{+1.9}$  &$4.9^{+5.3}_{-3.6}$ & $4013.1$ & $B^{\ast}\bar{B}^*$ & $1.0_{-0.1}^{+0.0}$  &$38.1^{+10.9}_{-10.5}$ & $10611.9$  \\
  \hline \hline
\end{tabular}
\end{table}

 The binding energies (if bound states exist) and scattering lengths of the $D^{(\ast)}\bar{D}^{(\ast)}$ system are  shown in Table \ref{tab:comparison-EFT1}. For the $D^{(\ast)}\bar{D}^{(\ast)}$ system, we obtain a $J^{PC}=2^{++}$~$D^{\ast}\bar{D}^{\ast}$ bound state with a binding energy of 5 MeV, namely $X(4013)$, which has already been predicted by many theoretical studies~\cite{Nieves:2012tt,Guo:2013sya,Baru:2016iwj}.  However, its existence  has not been confirmed by experiments. Except for $X(3872)$ and $X(4013)$ there are no other bound states in this system, which is consistent with the conclusion of the OBE model~\cite{Liu:2019stu}.  From the scattering length of $J^{PC}=0^{++}$~$\bar{D}D$ we find that the potential between $\bar{D}$ and $D$ is  attractive, but  is too weak to form a bound state.  Recently a Lattice QCD study obtained a  shallow bound state located 4 MeV below the $D\bar{D}$ mass threshold~\cite{Prelovsek:2020eiw}. If we use the OBE model to reproduce the 4 MeV  binding energy, the cutoff needed is about $\Lambda=1.4$ GeV, whose value is close to but larger than that obtained in our previous work to reproduce $X(3872)$, i.e., $\Lambda=1.01$ GeV. We note that
 several theoretical works have predicted the existence of a bound state in the  $D\bar{D}$ channel~\cite{Gamermann:2006nm,Sun:2011uh,Wang:2020elp,Ding:2020dio,Dai:2020yfu}.

 With heavy quark flavor symmetry  we can also  study the $B^{(\ast)}\bar{B}^{(\ast)}$ system. We find 5 bound states, $J^{PC}=0^{++}$~$\bar{B}B$,  $J^{PC}=1^{++}$~ $\bar{B}B^{\ast}$, $J^{PC}=1^{+-}$~$\bar{B}B^{\ast}$, $J^{PC}=1^{+-}$~$\bar{B}^{\ast}B^{\ast}$, and
$J^{PC}=2^{++}$~$\bar{B}^{\ast}B^{\ast}$, consistent with the results of the OBE model~\cite{Liu:2019stu}.
Among them we can regard the $J^{PC}=1^{++}$~$\bar{B}B^{\ast}$ and $J^{PC}=2^{++}$~$\bar{B}^{\ast}B^{\ast}$ states, referred to as $X_{b}$ and $X_{b}^{\prime}$, as the hidden bottom partners of $X(3872)$ and $X(4013)$, respectively.  We note that T\"ornqbist has already predicted $X(4013)$ with a binding energy 45 MeV in the one pion exchange model~\cite{Tornqvist:1993ng}, but its existence has not been confirmed by experiments.

\subsection{Meson-baryon system}

\begin{table}[!h]
\centering
\caption{ Scattering lengths ($a$ in fm), binding energies ($B$ in MeV, if bound states exist) and mass spectra ($M$ in MeV) of prospective isoscalar  heavy meson-baryon molecules.   The  $c$  and  $b$  subscript  denote  the  charm  and bottom sector, respectively. The uncertainties originate from the HADS breaking of the order $30\%$.
  }
\label{tab:comparison-EFT3}
\begin{tabular}{c|cc|ccc|c|ccc}
\hline\hline
molecule & $I$ & $J^{PC}$  & $a_{c}$ (fm)
& $B_{c}$ (MeV)
& $M_{c}$ (MeV ) &molecule   & $a_{b}$ (fm)
& $B_{b}$ (MeV)
& $M_{b}$ (MeV ) \\
  \hline
  $D\Xi_{cc}$ & $0$ & $1/2^{-}$ & $3.3^{-9.1}_{-1.5}$ & $2.0^{+7.3}_{\dag}$ & ? & $\bar{B}\Xi_{bb}$ & $1.0_{-0.1}^{+0.2}$ & $27.2^{+16.3}_{-14.7}$ & 15378.8  \\
  $D\Xi_{cc}^{\ast}$ & $0$ & $3/2^{-}$   &  $3.2^{-9.5}_{-1.4}$
  & $2.1^{+7.5}_{\dag}$ & ? & $\bar{B}\Xi_{bb}^{\ast}$     &  $1.0_{-0.1}^{+0.2}$
  & $27.2^{+16.3}_{-14.7}$ & 15402.8                  \\       \hline
  $D^{\ast}\Xi_{cc}$ & $0$ & $1/2^{-}$ & $1.8_{-0.5}^{+3.5}$   & $8.7^{+13.3}_{-8.1}$ & 5621.3 &$\bar{B}^{\ast}\Xi_{bb}$ &  $0.9_{-0.0}^{+0.1}$   & $40.0^{+21.0}_{-19.6}$ & 15412.0\\
   $D^{\ast}\Xi_{cc}$ & $0$ & $3/2^{-}$  &$4.7^{-8.0}_{-2.7}$   &  $0.9^{+5.3}_{\dag}$ & $?$   &  $\bar{B}^{\ast}\Xi_{bb}$ & $1.0^{-0.1}_{+0.3}$   & $21.3^{+13.9}_{-12.2}$ & $15430.7$\\
  \hline
  $D^{\ast}\Xi_{cc}^{\ast}$ & $0$ & $1/2^{-}$ &$-0.9_{-0.9}^{+0.4}$   & $\dag$ & $\dag$  & $\bar{B}^{\ast}\Xi_{bb}^{\ast}$  &$2.1^{+15.2}_{-0.6}$   & $1.9^{+4.0}_{-1.9}$ & $15474.0$ \\
  $D^{\ast}\Xi_{cc}^{\ast}$ & $0$ & $3/2^{-}$ & $22.8_{-22.0}^{-24.6}$   &$0^{+2.8}_{\dag}$ & $?$ &  $\bar{B}^{\ast}\Xi_{bb}^{\ast}$  &$1.1_{-0.1}^{+0.4}$   &$15.7^{+11.4}_{-10.8}$ & $15460.3$\\
 $D^{\ast}\Xi_{cc}^{\ast}$ & $0$ & $5/2^{-}$ &$1.6_{-0.4}^{+1.7}$   &$12.9^{+15.9}_{-10.9}$ & $5723.2$  & $\bar{B}^{\ast}\Xi_{bb}^{\ast}$   &$0.9_{-0.1}^{+0.1}$   &$46.6^{+23.3}_{-22.0}$ & $15429.4$ \\
  \hline \hline
\end{tabular}
\end{table}

Using HADS we can extend the $D^{(\ast)}\bar{D}^{(\ast)}$ system to the  $D^{(\ast)}\Xi_{cc}^{(\ast)}$ system as shown in Fig. \ref{cha}, where a HQSS multiplet of meson-baryon molecules can be expected.  The contact-range potentials of the  $D^{(\ast)}\Xi_{cc}^{(\ast)}$ system are expressed as combinations of $C_{a}$ and $C_{b}$, which can be related to the couplings arising in the $D^{(\ast)}\bar{D}^{(\ast)}$ system by HADS. Therefore,  we take the same input to calculate the binding energies and scattering lengths of the $D^{(\ast)}\Xi_{cc}^{(\ast)}$ system, and the results are given in  Table \ref{tab:comparison-EFT3}. Although most $D^{(\ast)}\Xi_{cc}^{(\ast)}$ systems are bound with the central values of $C_a$ and $C_b$,  they may become unbound once the breaking of HADS is taken into account. Only two systems,  $J^{P}=1/2^{-}$~$D^{\ast}\Xi_{cc}$ and  $J^{P}=5/2^{-}$~$D^{\ast}\Xi_{cc}^{\ast}$, always remain bound. Thus in this work we predict the existence of two triply charmed pentaquark molecules.  It is to be noted that in our previous study, we also predicted a  $J^{P}=1/2^{-}$~$D^{\ast}\Xi_{cc}$ bound state in the OBE model~\cite{Liu:2018bkx}, while in Ref.~\cite{Guo:2013xga} a
$J^{P}=5/2^{-}$~$D^{\ast}\Xi_{cc}^{\ast}$ bound state is predicted using a contact-range potential.  These two molecules can decay into $\Omega_{ccc}\omega$ via triangle diagrams, and we  expect that they can be detected in future experiments.

With heavy quark flavor symmetry we also study the $\bar{B}^{(\ast)}\Xi_{bb}^{(\ast)}$ system, and  find that  bound states exist in all the channels within uncertainties induced by heavy quark symmetry breaking, which indicates that we obtain a complete HQSS multiplet of hadronic molecules.   Comparing the results with those of the meson-meson  system, we find that the meson-baryon molecular states  are more bound, in agreement with naive expectations because of the smaller reduced masses in the latter system.

\begin{table}[!h]
\centering
\caption{ Scattering lengths ($a$ in fm), binding energies ($B$ in MeV, if bound states exist) and mass spectra ($M$ in MeV) of prospective isoscalar  heavy baryon-baryon molecules.  The  $c$  and  $b$  subscript  denote  the  charm  and bottom sector, respectively.  The uncertainties originate from the HADS breaking of the order $30\%$.
  }
\label{tab:comparison-EFT09}
\begin{tabular}{c|cc|ccc|c|ccc}
\hline\hline
molecule & $I$ & $J^{PC}$  & $a_{c}$ (fm)
& $B_{c}$ (MeV)
& $M_{c}$ (MeV )  & molecule    & $a_{b}$ (fm)
& $B_{b}$ (MeV)
& $M_{b}$ (MeV ) \\
  \hline
  $\bar{\Xi}_{cc}\Xi_{cc}$ & $0$ & $0^{-+}$ & $1.8_{-0.5}^{+3.4}$  & $6.3^{+9.4}_{-5.8}$ & 7235.7 & $\bar{\Xi}_{bb}\Xi_{bb}$  & $0.9_{-0.1}^{+0.1}$  & $29.3^{+15.1}_{-14.1}$ & 20224.7\\
  $\bar{\Xi}_{cc}\Xi_{cc}$ & $0$ & $1^{--}$ & $1.5^{+1.2}_{-0.3}$  & $11.5^{+12.9}_{-9.3}$ & 7230.4 &  $\bar{\Xi}_{bb}\Xi_{bb}$ &$0.9^{+0.1}_{-0.1}$  & $38.2^{+18.2}_{-17.3}$ & 20215.8\\
  \hline
  $\bar{\Xi}_{cc}^{\ast}\Xi_{cc}$+$\bar{\Xi}_{cc}\Xi_{cc}^{\ast}$ & $0$ & $1^{-+}$ & $1.2^{+0.4}_{-0.2}$  & $25.0^{+19.7}_{-16.4}$ & 7323.0 & $\bar{\Xi}_{bb}^{\ast}\Xi_{bb}+\bar{\Xi}_{bb}\Xi_{bb}^{\ast}$& $0.8^{+0.1}_{-0.0}$  & $56.5^{+24.4}_{-23.6}$ & 20221.5\\
  $\bar{\Xi}_{cc}^{\ast}\Xi_{cc}-\bar{\Xi}_{cc}\Xi_{cc}^{\ast}$ & $0$ & $1^{--}$& $1.4^{+1.2}_{-0.2}$   & $11.9^{+13.1}_{-9.5}$ & 7336.1 &  $\bar{\Xi}_{bb}^{\ast}\Xi_{bb}-\bar{\Xi}_{bb}\Xi_{bb}^{\ast}$ & $0.9^{+0.1}_{-0.1}$   & $38.2^{+18.2}_{-17.3}$ & 20245.9\\
  $\bar{\Xi}_{cc}^{\ast}\Xi_{cc}+\bar{\Xi}_{cc}\Xi_{cc}^{\ast}$ & $0$ & $2^{-+}$& $1.2^{+0.4}_{-0.2}$   & $25.0^{+19.7}_{-16.4}$ & 7323.0  &  $\bar{\Xi}_{bb}^{\ast}\Xi_{bb}+\bar{\Xi}_{bb}\Xi_{bb}^{\ast}$ &
    $0.8^{+0.1}_{-0.0}$  & $56.5^{+24.4}_{-23.6}$ & 20221.5\\
  $\bar{\Xi}_{cc}^{\ast}\Xi_{cc}-\bar{\Xi}_{cc}\Xi_{cc}^{\ast}$ & $0$ & $2^{--}$ & $-2.6^{+42.6}_{+1.7}$  & $\dag$ & $\dag$ & $\bar{\Xi}_{bb}^{\ast}\Xi_{bb}-\bar{\Xi}_{bb}\Xi_{bb}^{\ast}$  & $1.3^{+0.8}_{-0.2}$    & $5.7_{-4.2}^{+5.5}$ & 20272.3\\
  \hline
   $\bar{\Xi}_{cc}^{\ast}\Xi_{cc}^{\ast}$ & $0$ & $0^{-+}$ & $-2.7^{+24.8}_{+1.8}$ &  $\dag$  & $\dag$ & $\bar{\Xi}_{bb}^{\ast}\Xi_{bb}^{\ast}$& $1.3^{+0.8}_{-0.2}$   & $5.7_{-4.2}^{+5.5}$ & 20296.3   \\  $\bar{\Xi}_{cc}^{\ast}\Xi_{cc}^{\ast}$ & $0$ & $1^{--}$ & $8.0^{-10.3}_{-5.6}$ &  $0.2^{+2.8}_{\dag}$  & $?$ & $\bar{\Xi}_{bb}^{\ast}\Xi_{bb}^{\ast}$& $1.1^{+0.3}_{-0.1}$   & $12.8^{+8.8}_{-7.6}$ & 20289.2\\
   $\bar{\Xi}_{cc}^{\ast}\Xi_{cc}^{\ast}$ & $0$ & $2^{-+}$ & $1.7_{-0.4}^{+2.7}$ & $6.9^{+9.7}_{-6.2}$ & $7447.1$ &$\bar{\Xi}_{bb}^{\ast}\Xi_{bb}^{\ast}$ & $0.9^{+0.1}_{-0.1}$   & $29.4^{+15.0}_{-14.2}$ & 20272.6\\
  $\bar{\Xi}_{cc}^{\ast}\Xi_{cc}^{\ast}$ & $0$ & $3^{--}$ & $1.1^{+0.5}_{-0.1}$ & $25.6^{+19.7}_{-16.6}$ & $7428.4$ &$\bar{\Xi}_{bb}^{\ast}\Xi_{bb}^{\ast}$ & $0.8^{+0.1}_{-0.0}$   & $56.5^{+24.4}_{-23.5}$ & 20245.5\\
  \hline \hline
\end{tabular}
\end{table}

\subsection{Baryon-anti-baryon system}

In this section, we further extend the $D^{(\ast)}\Xi_{cc}^{(\ast)}$ system to the $\bar{\Xi}_{cc}^{(\ast)}\Xi_{cc}^{(\ast)}$ system as shown in Fig. \ref{cha}. This baryon-antibaryon system contains 10 channels related by HQSS.  The contact-range potentials of the $\bar{\Xi}_{cc}^{(\ast)}\Xi_{cc}^{(\ast)}$ system can also be written   in terms of $C_{a}$ and $C_{b}$ dictated by HQSS.
We compute the binding energies and scattering lengths of the $\bar{\Xi}_{cc}^{(\ast)}\Xi_{cc}^{(\ast)}$ system, and show the results in Table \ref{tab:comparison-EFT09}.
We find seven bound states, i.e.,  $J^{PC}=0^{-+}$~$\bar{\Xi}_{cc}\Xi_{cc}$, $J^{PC}=1^{--}$~$\bar{\Xi}_{cc}\Xi_{cc}$, $J^{PC}=1^{-+}$~$\bar{\Xi}_{cc}\Xi_{cc}^{\ast}$, $J^{PC}=1^{--}$~$\bar{\Xi}_{cc}\Xi_{cc}^{\ast}$,
$J^{PC}=2^{-+}$~$\bar{\Xi}_{cc}\Xi_{cc}^{\ast}$, $J^{PC}=2^{-+}$~$\bar{\Xi}_{cc}^{\ast}\Xi_{cc}^{\ast}$,
and $J^{PC}=3^{--}$~$\bar{\Xi}_{cc}^{\ast}\Xi_{cc}^{\ast}$. The $\bar{\Xi}_{cc}\Xi_{cc}$ system~\cite{Meng:2017fwb} and the $\bar{\Xi}_{cc}^{\ast}\Xi_{cc}^{\ast}$ system~\cite{Yang:2019rgw} have been studied in the OBE model, which show that all of these systems are bound.  Assuming heavy quark flavor symmetry, we further investigate their hidden bottom partners $\bar{\Xi}_{bb}^{(\ast)}\Xi_{bb}^{(\ast)}$, and all the states are found to be bound as shown  in Table~\ref{tab:comparison-EFT09}, suggesting the emergence of a complete multiplet of hadronic molecules.

\begin{figure}[!h]
 \center{\includegraphics[width=16cm]  {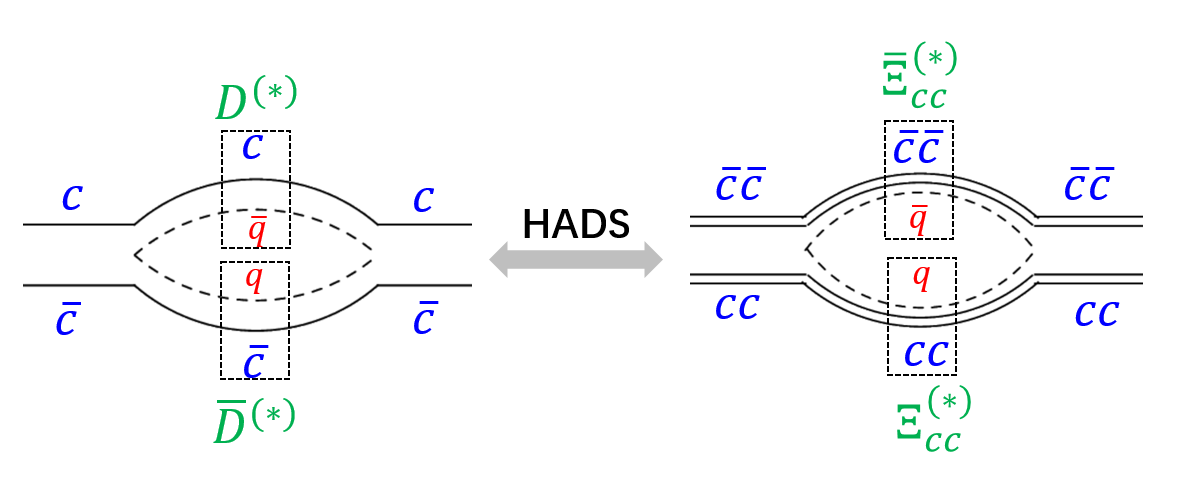}}
 \caption{\label{quark pair}Left:
 the two charmed meson channel heavily affects the mass spectra of charmonia close to their mass thresholds. Right: HADS implies that the two doubly charmed baryon channel could also affect the mass spectra of fully charmed tetraquark states close to their mass thresholds. }
 \end{figure}

Among these molecules, we note that there are two molecules, $J^{PC}=0^{-+}$~$\bar{\Xi}_{cc}\Xi_{cc}$ and $J^{PC}=1^{--}$~$\bar{\Xi}_{cc}\Xi_{cc}$ , located close to  7200 MeV,  which could contribute to the structure $X(7200)$ observed by the LHCb Collaboration recently.
These two molecules are predicted in terms of the hypothesis that $X(3872)$ is a $J^{PC}=1^{++}$~$\bar{D}D^{\ast}$ bound state and utilyzing the HADS.  It is well known that  the heavy meson-anti meson channel affects the  spectra of  conventional charmonia. According to the relationship  between $X(7200)$ and $X(3872)$ proposed by the present work  it is rather likely that the two doubly charmed baryon-antibaryon channel would affect the spectra of fully charmed tetraquark states  as shown in Fig.~\ref{quark pair}, which indicates that $X(7200)$ might contain both $\bar{\Xi}_{cc}\Xi_{cc}$ molecular and charmed diquark-anti diquark tetraquark components.  It is interesting to note that our study shows that
 $X(7200)$ might be a superposition of two molecular states.

\section{Summary}
\label{sum}
In this work we extended the $D^{(\ast)}\bar{D}^{(\ast)}$ system to the $D^{(\ast)}\Xi_{cc}^{(\ast)}$ system  then to the $\bar{\Xi}_{cc}^{(\ast)}\Xi_{cc}^{(\ast)}$ system utilyzing the heavy antiquark diquark symmetry (HADS).
The contact-range potentials of these systems respecting heavy quark spin symmetry (HQSS) are constrained to two couplings at the leading order, $C_{a}$ and $C_{b}$.
Assuming that $X(3872)$ is a $J^{PC}=1^{++}$~$\bar{D}D^{\ast}$ bound state, we determined the sum of $C_{a}$ and $C_{b}$.
To fully determine the value of $C_{a}$ and $C_{b}$, we used the light-meson saturation approach  to determine its ratio.
With the so-determined $C_{a}$ and $C_{b}$ we calculated the binding energies (if bound states exist) and scattering lengths of all these systems as well as their heavy quark flavor partners.

In the $\bar{D}^{(\ast)}D^{(\ast)}$ system, we obtained a  $J^{PC}=2^{++}$~$\bar{D}^{\ast}D^{\ast}$ molecule with a mass of $4013$ MeV referred to as $X(4013)$, which can be regarded as the heavy quark spin symmetry partner of $X(3872)$. Moreover, in the bottom sector we predicted the existence of five molecules. We extended the $\bar{D}^{(\ast)}D^{(\ast)}$ to the $D^{(\ast)}\Xi_{cc}^{(\ast)}$ system by HADS. Two molecules, $J^{P}=1/2^{-}$~$D^{\ast}\Xi_{cc}$ and $J^{P}=5/2^{-}$~$D^{\ast}\Xi_{cc}^{\ast}$, are obtained. There exist a complete multiplet of hadronic molecules in the corresponding bottom sector.    Using HADS again we studied the $\bar{\Xi}_{cc}^{(\ast)}{\Xi}_{cc}^{(\ast)}$ system, and found 7 bound states, $J^{PC}=0^{-+}$~$\bar{\Xi}_{cc}\Xi_{cc}$, $J^{PC}=1^{--}$~$\bar{\Xi}_{cc}\Xi_{cc}$, $J^{PC}=1^{-+}$~$\bar{\Xi}_{cc}\Xi_{cc}^{\ast}$, $J^{PC}=1^{--}$~$\bar{\Xi}_{cc}\Xi_{cc}^{\ast}$,
$J^{PC}=2^{-+}$~$\bar{\Xi}_{cc}\Xi_{cc}^{\ast}$, $J^{PC}=2^{-+}$~$\bar{\Xi}_{cc}^{\ast}\Xi_{cc}^{\ast}$
and $J^{PC}=3^{--}$~$\bar{\Xi}_{cc}^{\ast}\Xi_{cc}^{\ast}$. There also exist a complete multiplet of hadronic molecules in the bottom sector.
Among them, the $J^{PC}=0^{-+}$~$\bar{\Xi}_{cc}\Xi_{cc}$ and $J^{PC}=1^{--}$~$\bar{\Xi}_{cc}\Xi_{cc}$ molecules may contribute to $X(7200)$ recently observed by the LHCb Collaboration, which are related to $X(3872)$ via HADS.

\section{Acknowledgments}
 This work is partly supported by the National Natural Science Foundation of China under Grants No.11735003, No.11975041, and No.11961141004, and the fundamental Research Funds
for the Central Universities.

\bibliography{XiccSigmac}

\begin{thebibliography}{101}%
\makeatletter
\providecommand \@ifxundefined [1]{%
 \@ifx{#1\undefined}
}%
\providecommand \@ifnum [1]{%
 \ifnum #1\expandafter \@firstoftwo
 \else \expandafter \@secondoftwo
 \fi
}%
\providecommand \@ifx [1]{%
 \ifx #1\expandafter \@firstoftwo
 \else \expandafter \@secondoftwo
 \fi
}%
\providecommand \natexlab [1]{#1}%
\providecommand \enquote  [1]{``#1''}%
\providecommand \bibnamefont  [1]{#1}%
\providecommand \bibfnamefont [1]{#1}%
\providecommand \citenamefont [1]{#1}%
\providecommand \href@noop [0]{\@secondoftwo}%
\providecommand \href [0]{\begingroup \@sanitize@url \@href}%
\providecommand \@href[1]{\@@startlink{#1}\@@href}%
\providecommand \@@href[1]{\endgroup#1\@@endlink}%
\providecommand \@sanitize@url [0]{\catcode `\\12\catcode `\$12\catcode
  `\&12\catcode `\#12\catcode `\^12\catcode `\_12\catcode `\%12\relax}%
\providecommand \@@startlink[1]{}%
\providecommand \@@endlink[0]{}%
\providecommand \url  [0]{\begingroup\@sanitize@url \@url }%
\providecommand \@url [1]{\endgroup\@href {#1}{\urlprefix }}%
\providecommand \urlprefix  [0]{URL }%
\providecommand \Eprint [0]{\href }%
\providecommand \doibase [0]{http://dx.doi.org/}%
\providecommand \selectlanguage [0]{\@gobble}%
\providecommand \bibinfo  [0]{\@secondoftwo}%
\providecommand \bibfield  [0]{\@secondoftwo}%
\providecommand \translation [1]{[#1]}%
\providecommand \BibitemOpen [0]{}%
\providecommand \bibitemStop [0]{}%
\providecommand \bibitemNoStop [0]{.\EOS\space}%
\providecommand \EOS [0]{\spacefactor3000\relax}%
\providecommand \BibitemShut  [1]{\csname bibitem#1\endcsname}%
\let\auto@bib@innerbib\@empty
\bibitem [{\citenamefont {Gell-Mann}(1964)}]{GellMann:1964nj}%
  \BibitemOpen
  \bibfield  {author} {\bibinfo {author} {\bibfnamefont {M.}~\bibnamefont
  {Gell-Mann}},\ }\href {\doibase 10.1016/S0031-9163(64)92001-3} {\bibfield
  {journal} {\bibinfo  {journal} {Phys. Lett.}\ }\textbf {\bibinfo {volume}
  {8}},\ \bibinfo {pages} {214} (\bibinfo {year} {1964})}\BibitemShut {NoStop}%
\bibitem [{\citenamefont {Chen}\ \emph {et~al.}(2016)\citenamefont {Chen},
  \citenamefont {Chen}, \citenamefont {Liu},\ and\ \citenamefont
  {Zhu}}]{Chen:2016qju}%
  \BibitemOpen
  \bibfield  {author} {\bibinfo {author} {\bibfnamefont {H.-X.}\ \bibnamefont
  {Chen}}, \bibinfo {author} {\bibfnamefont {W.}~\bibnamefont {Chen}}, \bibinfo
  {author} {\bibfnamefont {X.}~\bibnamefont {Liu}}, \ and\ \bibinfo {author}
  {\bibfnamefont {S.-L.}\ \bibnamefont {Zhu}},\ }\href {\doibase
  10.1016/j.physrep.2016.05.004} {\bibfield  {journal} {\bibinfo  {journal}
  {Phys. Rept.}\ }\textbf {\bibinfo {volume} {639}},\ \bibinfo {pages} {1}
  (\bibinfo {year} {2016})},\ \Eprint {http://arxiv.org/abs/1601.02092}
  {arXiv:1601.02092 [hep-ph]} \BibitemShut {NoStop}%
\bibitem [{\citenamefont {Hosaka}\ \emph {et~al.}(2017)\citenamefont {Hosaka},
  \citenamefont {Hyodo}, \citenamefont {Sudoh}, \citenamefont {Yamaguchi},\
  and\ \citenamefont {Yasui}}]{Hosaka:2016ypm}%
  \BibitemOpen
  \bibfield  {author} {\bibinfo {author} {\bibfnamefont {A.}~\bibnamefont
  {Hosaka}}, \bibinfo {author} {\bibfnamefont {T.}~\bibnamefont {Hyodo}},
  \bibinfo {author} {\bibfnamefont {K.}~\bibnamefont {Sudoh}}, \bibinfo
  {author} {\bibfnamefont {Y.}~\bibnamefont {Yamaguchi}}, \ and\ \bibinfo
  {author} {\bibfnamefont {S.}~\bibnamefont {Yasui}},\ }\href {\doibase
  10.1016/j.ppnp.2017.04.003} {\bibfield  {journal} {\bibinfo  {journal} {Prog.
  Part. Nucl. Phys.}\ }\textbf {\bibinfo {volume} {96}},\ \bibinfo {pages} {88}
  (\bibinfo {year} {2017})},\ \Eprint {http://arxiv.org/abs/1606.08685}
  {arXiv:1606.08685 [hep-ph]} \BibitemShut {NoStop}%
\bibitem [{\citenamefont {Guo}\ \emph {et~al.}(2018)\citenamefont {Guo},
  \citenamefont {Hanhart}, \citenamefont {Mei\ss{}ner}, \citenamefont {Wang},
  \citenamefont {Zhao},\ and\ \citenamefont {Zou}}]{Guo:2017jvc}%
  \BibitemOpen
  \bibfield  {author} {\bibinfo {author} {\bibfnamefont {F.-K.}\ \bibnamefont
  {Guo}}, \bibinfo {author} {\bibfnamefont {C.}~\bibnamefont {Hanhart}},
  \bibinfo {author} {\bibfnamefont {U.-G.}\ \bibnamefont {Mei\ss{}ner}},
  \bibinfo {author} {\bibfnamefont {Q.}~\bibnamefont {Wang}}, \bibinfo {author}
  {\bibfnamefont {Q.}~\bibnamefont {Zhao}}, \ and\ \bibinfo {author}
  {\bibfnamefont {B.-S.}\ \bibnamefont {Zou}},\ }\href {\doibase
  10.1103/RevModPhys.90.015004} {\bibfield  {journal} {\bibinfo  {journal}
  {Rev. Mod. Phys.}\ }\textbf {\bibinfo {volume} {90}},\ \bibinfo {pages}
  {015004} (\bibinfo {year} {2018})},\ \Eprint
  {http://arxiv.org/abs/1705.00141} {arXiv:1705.00141 [hep-ph]} \BibitemShut
  {NoStop}%
\bibitem [{\citenamefont {Olsen}\ \emph {et~al.}(2018)\citenamefont {Olsen},
  \citenamefont {Skwarnicki},\ and\ \citenamefont {Zieminska}}]{Olsen:2017bmm}%
  \BibitemOpen
  \bibfield  {author} {\bibinfo {author} {\bibfnamefont {S.~L.}\ \bibnamefont
  {Olsen}}, \bibinfo {author} {\bibfnamefont {T.}~\bibnamefont {Skwarnicki}}, \
  and\ \bibinfo {author} {\bibfnamefont {D.}~\bibnamefont {Zieminska}},\ }\href
  {\doibase 10.1103/RevModPhys.90.015003} {\bibfield  {journal} {\bibinfo
  {journal} {Rev. Mod. Phys.}\ }\textbf {\bibinfo {volume} {90}},\ \bibinfo
  {pages} {015003} (\bibinfo {year} {2018})},\ \Eprint
  {http://arxiv.org/abs/1708.04012} {arXiv:1708.04012 [hep-ph]} \BibitemShut
  {NoStop}%
\bibitem [{\citenamefont {Brambilla}\ \emph {et~al.}(2020)\citenamefont
  {Brambilla}, \citenamefont {Eidelman}, \citenamefont {Hanhart}, \citenamefont
  {Nefediev}, \citenamefont {Shen}, \citenamefont {Thomas}, \citenamefont
  {Vairo},\ and\ \citenamefont {Yuan}}]{Brambilla:2019esw}%
  \BibitemOpen
  \bibfield  {author} {\bibinfo {author} {\bibfnamefont {N.}~\bibnamefont
  {Brambilla}}, \bibinfo {author} {\bibfnamefont {S.}~\bibnamefont {Eidelman}},
  \bibinfo {author} {\bibfnamefont {C.}~\bibnamefont {Hanhart}}, \bibinfo
  {author} {\bibfnamefont {A.}~\bibnamefont {Nefediev}}, \bibinfo {author}
  {\bibfnamefont {C.-P.}\ \bibnamefont {Shen}}, \bibinfo {author}
  {\bibfnamefont {C.~E.}\ \bibnamefont {Thomas}}, \bibinfo {author}
  {\bibfnamefont {A.}~\bibnamefont {Vairo}}, \ and\ \bibinfo {author}
  {\bibfnamefont {C.-Z.}\ \bibnamefont {Yuan}},\ }\href@noop {} {\bibfield
  {journal} {\bibinfo  {journal} {Phys. Rept.}\ }\textbf {\bibinfo {volume}
  {873}},\ \bibinfo {pages} {1} (\bibinfo {year} {2020})},\ \Eprint
  {http://arxiv.org/abs/1907.07583} {arXiv:1907.07583 [hep-ex]} \BibitemShut
  {NoStop}%
\bibitem [{\citenamefont {Aaij}\ \emph {et~al.}(2015)\citenamefont {Aaij} \emph
  {et~al.}}]{Aaij:2015tga}%
  \BibitemOpen
  \bibfield  {author} {\bibinfo {author} {\bibfnamefont {R.}~\bibnamefont
  {Aaij}} \emph {et~al.} (\bibinfo {collaboration} {LHCb}),\ }\href {\doibase
  10.1103/PhysRevLett.115.072001} {\bibfield  {journal} {\bibinfo  {journal}
  {Phys. Rev. Lett.}\ }\textbf {\bibinfo {volume} {115}},\ \bibinfo {pages}
  {072001} (\bibinfo {year} {2015})},\ \Eprint
  {http://arxiv.org/abs/1507.03414} {arXiv:1507.03414 [hep-ex]} \BibitemShut
  {NoStop}%
\bibitem [{\citenamefont {Aubert}\ \emph
  {et~al.}(2005{\natexlab{a}})\citenamefont {Aubert} \emph
  {et~al.}}]{Aubert:2005rm}%
  \BibitemOpen
  \bibfield  {author} {\bibinfo {author} {\bibfnamefont {B.}~\bibnamefont
  {Aubert}} \emph {et~al.} (\bibinfo {collaboration} {BaBar}),\ }\href
  {\doibase 10.1103/PhysRevLett.95.142001} {\bibfield  {journal} {\bibinfo
  {journal} {Phys. Rev. Lett.}\ }\textbf {\bibinfo {volume} {95}},\ \bibinfo
  {pages} {142001} (\bibinfo {year} {2005}{\natexlab{a}})},\ \Eprint
  {http://arxiv.org/abs/hep-ex/0506081} {arXiv:hep-ex/0506081 [hep-ex]}
  \BibitemShut {NoStop}%
\bibitem [{\citenamefont {Acosta}\ \emph {et~al.}(2004)\citenamefont {Acosta}
  \emph {et~al.}}]{Acosta:2003zx}%
  \BibitemOpen
  \bibfield  {author} {\bibinfo {author} {\bibfnamefont {D.}~\bibnamefont
  {Acosta}} \emph {et~al.} (\bibinfo {collaboration} {CDF}),\ }\href {\doibase
  10.1103/PhysRevLett.93.072001} {\bibfield  {journal} {\bibinfo  {journal}
  {Phys. Rev. Lett.}\ }\textbf {\bibinfo {volume} {93}},\ \bibinfo {pages}
  {072001} (\bibinfo {year} {2004})},\ \Eprint
  {http://arxiv.org/abs/hep-ex/0312021} {arXiv:hep-ex/0312021 [hep-ex]}
  \BibitemShut {NoStop}%
\bibitem [{\citenamefont {Abazov}\ \emph {et~al.}(2004)\citenamefont {Abazov}
  \emph {et~al.}}]{Abazov:2004kp}%
  \BibitemOpen
  \bibfield  {author} {\bibinfo {author} {\bibfnamefont {V.~M.}\ \bibnamefont
  {Abazov}} \emph {et~al.} (\bibinfo {collaboration} {D0}),\ }\href {\doibase
  10.1103/PhysRevLett.93.162002} {\bibfield  {journal} {\bibinfo  {journal}
  {Phys. Rev. Lett.}\ }\textbf {\bibinfo {volume} {93}},\ \bibinfo {pages}
  {162002} (\bibinfo {year} {2004})},\ \Eprint
  {http://arxiv.org/abs/hep-ex/0405004} {arXiv:hep-ex/0405004 [hep-ex]}
  \BibitemShut {NoStop}%
\bibitem [{\citenamefont {Aubert}\ \emph
  {et~al.}(2005{\natexlab{b}})\citenamefont {Aubert} \emph
  {et~al.}}]{Aubert:2004ns}%
  \BibitemOpen
  \bibfield  {author} {\bibinfo {author} {\bibfnamefont {B.}~\bibnamefont
  {Aubert}} \emph {et~al.} (\bibinfo {collaboration} {BaBar}),\ }\href
  {\doibase 10.1103/PhysRevD.71.071103} {\bibfield  {journal} {\bibinfo
  {journal} {Phys. Rev.}\ }\textbf {\bibinfo {volume} {D71}},\ \bibinfo {pages}
  {071103} (\bibinfo {year} {2005}{\natexlab{b}})},\ \Eprint
  {http://arxiv.org/abs/hep-ex/0406022} {arXiv:hep-ex/0406022 [hep-ex]}
  \BibitemShut {NoStop}%
\bibitem [{\citenamefont {Ablikim}\ \emph {et~al.}(2014)\citenamefont {Ablikim}
  \emph {et~al.}}]{Ablikim:2013dyn}%
  \BibitemOpen
  \bibfield  {author} {\bibinfo {author} {\bibfnamefont {M.}~\bibnamefont
  {Ablikim}} \emph {et~al.} (\bibinfo {collaboration} {BESIII}),\ }\href
  {\doibase 10.1103/PhysRevLett.112.092001} {\bibfield  {journal} {\bibinfo
  {journal} {Phys. Rev. Lett.}\ }\textbf {\bibinfo {volume} {112}},\ \bibinfo
  {pages} {092001} (\bibinfo {year} {2014})},\ \Eprint
  {http://arxiv.org/abs/1310.4101} {arXiv:1310.4101 [hep-ex]} \BibitemShut
  {NoStop}%
\bibitem [{\citenamefont {Aaij}\ \emph {et~al.}(2012)\citenamefont {Aaij} \emph
  {et~al.}}]{Aaij:2011sn}%
  \BibitemOpen
  \bibfield  {author} {\bibinfo {author} {\bibfnamefont {R.}~\bibnamefont
  {Aaij}} \emph {et~al.} (\bibinfo {collaboration} {LHCb}),\ }\href {\doibase
  10.1140/epjc/s10052-012-1972-7} {\bibfield  {journal} {\bibinfo  {journal}
  {Eur. Phys. J.}\ }\textbf {\bibinfo {volume} {C72}},\ \bibinfo {pages} {1972}
  (\bibinfo {year} {2012})},\ \Eprint {http://arxiv.org/abs/1112.5310}
  {arXiv:1112.5310 [hep-ex]} \BibitemShut {NoStop}%
\bibitem [{\citenamefont {Aaij}\ \emph {et~al.}(2013)\citenamefont {Aaij} \emph
  {et~al.}}]{Aaij:2013zoa}%
  \BibitemOpen
  \bibfield  {author} {\bibinfo {author} {\bibfnamefont {R.}~\bibnamefont
  {Aaij}} \emph {et~al.} (\bibinfo {collaboration} {LHCb}),\ }\href {\doibase
  10.1103/PhysRevLett.110.222001} {\bibfield  {journal} {\bibinfo  {journal}
  {Phys. Rev. Lett.}\ }\textbf {\bibinfo {volume} {110}},\ \bibinfo {pages}
  {222001} (\bibinfo {year} {2013})},\ \Eprint {http://arxiv.org/abs/1302.6269}
  {arXiv:1302.6269 [hep-ex]} \BibitemShut {NoStop}%
\bibitem [{\citenamefont {Abe}\ \emph {et~al.}(2005)\citenamefont {Abe} \emph
  {et~al.}}]{Abe:2005ix}%
  \BibitemOpen
  \bibfield  {author} {\bibinfo {author} {\bibfnamefont {K.}~\bibnamefont
  {Abe}} \emph {et~al.} (\bibinfo {collaboration} {Belle}),\ }in\ \href@noop {}
  {\emph {\bibinfo {booktitle} {{Lepton and photon interactions at high
  energies. Proceedings, 22nd International Symposium, LP 2005, Uppsala,
  Sweden, June 30-July 5, 2005}}}}\ (\bibinfo {year} {2005})\ \Eprint
  {http://arxiv.org/abs/hep-ex/0505037} {arXiv:hep-ex/0505037 [hep-ex]}
  \BibitemShut {NoStop}%
\bibitem [{\citenamefont {del Amo~Sanchez}\ \emph {et~al.}(2010)\citenamefont
  {del Amo~Sanchez} \emph {et~al.}}]{delAmoSanchez:2010jr}%
  \BibitemOpen
  \bibfield  {author} {\bibinfo {author} {\bibfnamefont {P.}~\bibnamefont {del
  Amo~Sanchez}} \emph {et~al.} (\bibinfo {collaboration} {BaBar}),\ }\href
  {\doibase 10.1103/PhysRevD.82.011101} {\bibfield  {journal} {\bibinfo
  {journal} {Phys. Rev.}\ }\textbf {\bibinfo {volume} {D82}},\ \bibinfo {pages}
  {011101} (\bibinfo {year} {2010})},\ \Eprint {http://arxiv.org/abs/1005.5190}
  {arXiv:1005.5190 [hep-ex]} \BibitemShut {NoStop}%
\bibitem [{\citenamefont {Swanson}(2004)}]{Swanson:2003tb}%
  \BibitemOpen
  \bibfield  {author} {\bibinfo {author} {\bibfnamefont {E.~S.}\ \bibnamefont
  {Swanson}},\ }\href {\doibase 10.1016/j.physletb.2004.03.033} {\bibfield
  {journal} {\bibinfo  {journal} {Phys. Lett. B}\ }\textbf {\bibinfo {volume}
  {588}},\ \bibinfo {pages} {189} (\bibinfo {year} {2004})},\ \Eprint
  {http://arxiv.org/abs/hep-ph/0311229} {arXiv:hep-ph/0311229} \BibitemShut
  {NoStop}%
\bibitem [{\citenamefont {Voloshin}(2004)}]{Voloshin:2003nt}%
  \BibitemOpen
  \bibfield  {author} {\bibinfo {author} {\bibfnamefont {M.}~\bibnamefont
  {Voloshin}},\ }\href {\doibase 10.1016/j.physletb.2003.11.014} {\bibfield
  {journal} {\bibinfo  {journal} {Phys. Lett. B}\ }\textbf {\bibinfo {volume}
  {579}},\ \bibinfo {pages} {316} (\bibinfo {year} {2004})},\ \Eprint
  {http://arxiv.org/abs/hep-ph/0309307} {arXiv:hep-ph/0309307} \BibitemShut
  {NoStop}%
\bibitem [{\citenamefont {AlFiky}\ \emph {et~al.}(2006)\citenamefont {AlFiky},
  \citenamefont {Gabbiani},\ and\ \citenamefont {Petrov}}]{AlFiky:2005jd}%
  \BibitemOpen
  \bibfield  {author} {\bibinfo {author} {\bibfnamefont {M.~T.}\ \bibnamefont
  {AlFiky}}, \bibinfo {author} {\bibfnamefont {F.}~\bibnamefont {Gabbiani}}, \
  and\ \bibinfo {author} {\bibfnamefont {A.~A.}\ \bibnamefont {Petrov}},\
  }\href {\doibase 10.1016/j.physletb.2006.07.069} {\bibfield  {journal}
  {\bibinfo  {journal} {Phys. Lett. B}\ }\textbf {\bibinfo {volume} {640}},\
  \bibinfo {pages} {238} (\bibinfo {year} {2006})},\ \Eprint
  {http://arxiv.org/abs/hep-ph/0506141} {arXiv:hep-ph/0506141} \BibitemShut
  {NoStop}%
\bibitem [{\citenamefont {Liu}\ \emph {et~al.}(2008)\citenamefont {Liu},
  \citenamefont {Liu}, \citenamefont {Deng},\ and\ \citenamefont
  {Zhu}}]{Liu:2008fh}%
  \BibitemOpen
  \bibfield  {author} {\bibinfo {author} {\bibfnamefont {Y.-R.}\ \bibnamefont
  {Liu}}, \bibinfo {author} {\bibfnamefont {X.}~\bibnamefont {Liu}}, \bibinfo
  {author} {\bibfnamefont {W.-Z.}\ \bibnamefont {Deng}}, \ and\ \bibinfo
  {author} {\bibfnamefont {S.-L.}\ \bibnamefont {Zhu}},\ }\href {\doibase
  10.1140/epjc/s10052-008-0640-4} {\bibfield  {journal} {\bibinfo  {journal}
  {Eur. Phys. J.}\ }\textbf {\bibinfo {volume} {C56}},\ \bibinfo {pages} {63}
  (\bibinfo {year} {2008})},\ \Eprint {http://arxiv.org/abs/0801.3540}
  {arXiv:0801.3540 [hep-ph]} \BibitemShut {NoStop}%
\bibitem [{\citenamefont {Sun}\ \emph {et~al.}(2011)\citenamefont {Sun},
  \citenamefont {He}, \citenamefont {Liu}, \citenamefont {Luo},\ and\
  \citenamefont {Zhu}}]{Sun:2011uh}%
  \BibitemOpen
  \bibfield  {author} {\bibinfo {author} {\bibfnamefont {Z.-F.}\ \bibnamefont
  {Sun}}, \bibinfo {author} {\bibfnamefont {J.}~\bibnamefont {He}}, \bibinfo
  {author} {\bibfnamefont {X.}~\bibnamefont {Liu}}, \bibinfo {author}
  {\bibfnamefont {Z.-G.}\ \bibnamefont {Luo}}, \ and\ \bibinfo {author}
  {\bibfnamefont {S.-L.}\ \bibnamefont {Zhu}},\ }\href {\doibase
  10.1103/PhysRevD.84.054002} {\bibfield  {journal} {\bibinfo  {journal} {Phys.
  Rev. D}\ }\textbf {\bibinfo {volume} {84}},\ \bibinfo {pages} {054002}
  (\bibinfo {year} {2011})},\ \Eprint {http://arxiv.org/abs/1106.2968}
  {arXiv:1106.2968 [hep-ph]} \BibitemShut {NoStop}%
\bibitem [{\citenamefont {Nieves}\ and\ \citenamefont
  {Valderrama}(2012)}]{Nieves:2012tt}%
  \BibitemOpen
  \bibfield  {author} {\bibinfo {author} {\bibfnamefont {J.}~\bibnamefont
  {Nieves}}\ and\ \bibinfo {author} {\bibfnamefont {M.~P.}\ \bibnamefont
  {Valderrama}},\ }\href {\doibase 10.1103/PhysRevD.86.056004} {\bibfield
  {journal} {\bibinfo  {journal} {Phys. Rev.}\ }\textbf {\bibinfo {volume}
  {D86}},\ \bibinfo {pages} {056004} (\bibinfo {year} {2012})},\ \Eprint
  {http://arxiv.org/abs/1204.2790} {arXiv:1204.2790 [hep-ph]} \BibitemShut
  {NoStop}%
\bibitem [{\citenamefont {Guo}\ \emph {et~al.}(2013{\natexlab{a}})\citenamefont
  {Guo}, \citenamefont {Hidalgo-Duque}, \citenamefont {Nieves},\ and\
  \citenamefont {Valderrama}}]{Guo:2013sya}%
  \BibitemOpen
  \bibfield  {author} {\bibinfo {author} {\bibfnamefont {F.-K.}\ \bibnamefont
  {Guo}}, \bibinfo {author} {\bibfnamefont {C.}~\bibnamefont {Hidalgo-Duque}},
  \bibinfo {author} {\bibfnamefont {J.}~\bibnamefont {Nieves}}, \ and\ \bibinfo
  {author} {\bibfnamefont {M.~P.}\ \bibnamefont {Valderrama}},\ }\href
  {\doibase 10.1103/PhysRevD.88.054007} {\bibfield  {journal} {\bibinfo
  {journal} {Phys. Rev. D}\ }\textbf {\bibinfo {volume} {88}},\ \bibinfo
  {pages} {054007} (\bibinfo {year} {2013}{\natexlab{a}})},\ \Eprint
  {http://arxiv.org/abs/1303.6608} {arXiv:1303.6608 [hep-ph]} \BibitemShut
  {NoStop}%
\bibitem [{\citenamefont {Karliner}\ and\ \citenamefont
  {Rosner}(2015)}]{Karliner:2015ina}%
  \BibitemOpen
  \bibfield  {author} {\bibinfo {author} {\bibfnamefont {M.}~\bibnamefont
  {Karliner}}\ and\ \bibinfo {author} {\bibfnamefont {J.~L.}\ \bibnamefont
  {Rosner}},\ }\href {\doibase 10.1103/PhysRevLett.115.122001} {\bibfield
  {journal} {\bibinfo  {journal} {Phys. Rev. Lett.}\ }\textbf {\bibinfo
  {volume} {115}},\ \bibinfo {pages} {122001} (\bibinfo {year} {2015})},\
  \Eprint {http://arxiv.org/abs/1506.06386} {arXiv:1506.06386 [hep-ph]}
  \BibitemShut {NoStop}%
\bibitem [{\citenamefont {Liu}\ \emph {et~al.}(2019{\natexlab{a}})\citenamefont
  {Liu}, \citenamefont {Wu}, \citenamefont {Pavon~Valderrama}, \citenamefont
  {Xie},\ and\ \citenamefont {Geng}}]{Liu:2019stu}%
  \BibitemOpen
  \bibfield  {author} {\bibinfo {author} {\bibfnamefont {M.-Z.}\ \bibnamefont
  {Liu}}, \bibinfo {author} {\bibfnamefont {T.-W.}\ \bibnamefont {Wu}},
  \bibinfo {author} {\bibfnamefont {M.}~\bibnamefont {Pavon~Valderrama}},
  \bibinfo {author} {\bibfnamefont {J.-J.}\ \bibnamefont {Xie}}, \ and\
  \bibinfo {author} {\bibfnamefont {L.-S.}\ \bibnamefont {Geng}},\ }\href
  {\doibase 10.1103/PhysRevD.99.094018} {\bibfield  {journal} {\bibinfo
  {journal} {Phys. Rev. D}\ }\textbf {\bibinfo {volume} {99}},\ \bibinfo
  {pages} {094018} (\bibinfo {year} {2019}{\natexlab{a}})},\ \Eprint
  {http://arxiv.org/abs/1902.03044} {arXiv:1902.03044 [hep-ph]} \BibitemShut
  {NoStop}%
\bibitem [{\citenamefont {Maiani}\ \emph {et~al.}(2005)\citenamefont {Maiani},
  \citenamefont {Piccinini}, \citenamefont {Polosa},\ and\ \citenamefont
  {Riquer}}]{Maiani:2004vq}%
  \BibitemOpen
  \bibfield  {author} {\bibinfo {author} {\bibfnamefont {L.}~\bibnamefont
  {Maiani}}, \bibinfo {author} {\bibfnamefont {F.}~\bibnamefont {Piccinini}},
  \bibinfo {author} {\bibfnamefont {A.}~\bibnamefont {Polosa}}, \ and\ \bibinfo
  {author} {\bibfnamefont {V.}~\bibnamefont {Riquer}},\ }\href {\doibase
  10.1103/PhysRevD.71.014028} {\bibfield  {journal} {\bibinfo  {journal} {Phys.
  Rev. D}\ }\textbf {\bibinfo {volume} {71}},\ \bibinfo {pages} {014028}
  (\bibinfo {year} {2005})},\ \Eprint {http://arxiv.org/abs/hep-ph/0412098}
  {arXiv:hep-ph/0412098} \BibitemShut {NoStop}%
\bibitem [{\citenamefont {Ebert}\ \emph {et~al.}(2006)\citenamefont {Ebert},
  \citenamefont {Faustov},\ and\ \citenamefont {Galkin}}]{Ebert:2005nc}%
  \BibitemOpen
  \bibfield  {author} {\bibinfo {author} {\bibfnamefont {D.}~\bibnamefont
  {Ebert}}, \bibinfo {author} {\bibfnamefont {R.}~\bibnamefont {Faustov}}, \
  and\ \bibinfo {author} {\bibfnamefont {V.}~\bibnamefont {Galkin}},\ }\href
  {\doibase 10.1016/j.physletb.2006.01.026} {\bibfield  {journal} {\bibinfo
  {journal} {Phys. Lett. B}\ }\textbf {\bibinfo {volume} {634}},\ \bibinfo
  {pages} {214} (\bibinfo {year} {2006})},\ \Eprint
  {http://arxiv.org/abs/hep-ph/0512230} {arXiv:hep-ph/0512230} \BibitemShut
  {NoStop}%
\bibitem [{\citenamefont {Matheus}\ \emph {et~al.}(2007)\citenamefont
  {Matheus}, \citenamefont {Narison}, \citenamefont {Nielsen},\ and\
  \citenamefont {Richard}}]{Matheus:2006xi}%
  \BibitemOpen
  \bibfield  {author} {\bibinfo {author} {\bibfnamefont {R.~D.}\ \bibnamefont
  {Matheus}}, \bibinfo {author} {\bibfnamefont {S.}~\bibnamefont {Narison}},
  \bibinfo {author} {\bibfnamefont {M.}~\bibnamefont {Nielsen}}, \ and\
  \bibinfo {author} {\bibfnamefont {J.}~\bibnamefont {Richard}},\ }\href
  {\doibase 10.1103/PhysRevD.75.014005} {\bibfield  {journal} {\bibinfo
  {journal} {Phys. Rev. D}\ }\textbf {\bibinfo {volume} {75}},\ \bibinfo
  {pages} {014005} (\bibinfo {year} {2007})},\ \Eprint
  {http://arxiv.org/abs/hep-ph/0608297} {arXiv:hep-ph/0608297} \BibitemShut
  {NoStop}%
\bibitem [{\citenamefont {Wang}\ and\ \citenamefont
  {Huang}(2014)}]{Wang:2013vex}%
  \BibitemOpen
  \bibfield  {author} {\bibinfo {author} {\bibfnamefont {Z.-G.}\ \bibnamefont
  {Wang}}\ and\ \bibinfo {author} {\bibfnamefont {T.}~\bibnamefont {Huang}},\
  }\href {\doibase 10.1103/PhysRevD.89.054019} {\bibfield  {journal} {\bibinfo
  {journal} {Phys. Rev. D}\ }\textbf {\bibinfo {volume} {89}},\ \bibinfo
  {pages} {054019} (\bibinfo {year} {2014})},\ \Eprint
  {http://arxiv.org/abs/1310.2422} {arXiv:1310.2422 [hep-ph]} \BibitemShut
  {NoStop}%
\bibitem [{\citenamefont {Barnes}\ and\ \citenamefont
  {Godfrey}(2004)}]{Barnes:2003vb}%
  \BibitemOpen
  \bibfield  {author} {\bibinfo {author} {\bibfnamefont {T.}~\bibnamefont
  {Barnes}}\ and\ \bibinfo {author} {\bibfnamefont {S.}~\bibnamefont
  {Godfrey}},\ }\href {\doibase 10.1103/PhysRevD.69.054008} {\bibfield
  {journal} {\bibinfo  {journal} {Phys. Rev. D}\ }\textbf {\bibinfo {volume}
  {69}},\ \bibinfo {pages} {054008} (\bibinfo {year} {2004})},\ \Eprint
  {http://arxiv.org/abs/hep-ph/0311162} {arXiv:hep-ph/0311162} \BibitemShut
  {NoStop}%
\bibitem [{\citenamefont {Vijande}\ \emph {et~al.}(2005)\citenamefont
  {Vijande}, \citenamefont {Fernandez},\ and\ \citenamefont
  {Valcarce}}]{Vijande:2004he}%
  \BibitemOpen
  \bibfield  {author} {\bibinfo {author} {\bibfnamefont {J.}~\bibnamefont
  {Vijande}}, \bibinfo {author} {\bibfnamefont {F.}~\bibnamefont {Fernandez}},
  \ and\ \bibinfo {author} {\bibfnamefont {A.}~\bibnamefont {Valcarce}},\
  }\href {\doibase 10.1088/0954-3899/31/5/017} {\bibfield  {journal} {\bibinfo
  {journal} {J. Phys. G}\ }\textbf {\bibinfo {volume} {31}},\ \bibinfo {pages}
  {481} (\bibinfo {year} {2005})},\ \Eprint
  {http://arxiv.org/abs/hep-ph/0411299} {arXiv:hep-ph/0411299} \BibitemShut
  {NoStop}%
\bibitem [{\citenamefont {Kalashnikova}(2005)}]{Kalashnikova:2005ui}%
  \BibitemOpen
  \bibfield  {author} {\bibinfo {author} {\bibfnamefont {Y.}~\bibnamefont
  {Kalashnikova}},\ }\href {\doibase 10.1103/PhysRevD.72.034010} {\bibfield
  {journal} {\bibinfo  {journal} {Phys. Rev. D}\ }\textbf {\bibinfo {volume}
  {72}},\ \bibinfo {pages} {034010} (\bibinfo {year} {2005})},\ \Eprint
  {http://arxiv.org/abs/hep-ph/0506270} {arXiv:hep-ph/0506270} \BibitemShut
  {NoStop}%
\bibitem [{\citenamefont {Matheus}\ \emph {et~al.}(2009)\citenamefont
  {Matheus}, \citenamefont {Navarra}, \citenamefont {Nielsen},\ and\
  \citenamefont {Zanetti}}]{Matheus:2009vq}%
  \BibitemOpen
  \bibfield  {author} {\bibinfo {author} {\bibfnamefont {R.~D.}\ \bibnamefont
  {Matheus}}, \bibinfo {author} {\bibfnamefont {F.}~\bibnamefont {Navarra}},
  \bibinfo {author} {\bibfnamefont {M.}~\bibnamefont {Nielsen}}, \ and\
  \bibinfo {author} {\bibfnamefont {C.}~\bibnamefont {Zanetti}},\ }\href
  {\doibase 10.1103/PhysRevD.80.056002} {\bibfield  {journal} {\bibinfo
  {journal} {Phys. Rev. D}\ }\textbf {\bibinfo {volume} {80}},\ \bibinfo
  {pages} {056002} (\bibinfo {year} {2009})},\ \Eprint
  {http://arxiv.org/abs/0907.2683} {arXiv:0907.2683 [hep-ph]} \BibitemShut
  {NoStop}%
\bibitem [{\citenamefont {Ferretti}\ \emph {et~al.}(2013)\citenamefont
  {Ferretti}, \citenamefont {Galat\`a},\ and\ \citenamefont
  {Santopinto}}]{Ferretti:2013faa}%
  \BibitemOpen
  \bibfield  {author} {\bibinfo {author} {\bibfnamefont {J.}~\bibnamefont
  {Ferretti}}, \bibinfo {author} {\bibfnamefont {G.}~\bibnamefont {Galat\`a}},
  \ and\ \bibinfo {author} {\bibfnamefont {E.}~\bibnamefont {Santopinto}},\
  }\href {\doibase 10.1103/PhysRevC.88.015207} {\bibfield  {journal} {\bibinfo
  {journal} {Phys. Rev. C}\ }\textbf {\bibinfo {volume} {88}},\ \bibinfo
  {pages} {015207} (\bibinfo {year} {2013})},\ \Eprint
  {http://arxiv.org/abs/1302.6857} {arXiv:1302.6857 [hep-ph]} \BibitemShut
  {NoStop}%
\bibitem [{\citenamefont {Barnes}\ \emph {et~al.}(2003)\citenamefont {Barnes},
  \citenamefont {Close},\ and\ \citenamefont {Lipkin}}]{Barnes:2003dj}%
  \BibitemOpen
  \bibfield  {author} {\bibinfo {author} {\bibfnamefont {T.}~\bibnamefont
  {Barnes}}, \bibinfo {author} {\bibfnamefont {F.}~\bibnamefont {Close}}, \
  and\ \bibinfo {author} {\bibfnamefont {H.}~\bibnamefont {Lipkin}},\ }\href
  {\doibase 10.1103/PhysRevD.68.054006} {\bibfield  {journal} {\bibinfo
  {journal} {Phys. Rev. D}\ }\textbf {\bibinfo {volume} {68}},\ \bibinfo
  {pages} {054006} (\bibinfo {year} {2003})},\ \Eprint
  {http://arxiv.org/abs/hep-ph/0305025} {arXiv:hep-ph/0305025} \BibitemShut
  {NoStop}%
\bibitem [{\citenamefont {Kolomeitsev}\ and\ \citenamefont
  {Lutz}(2004)}]{Kolomeitsev:2003ac}%
  \BibitemOpen
  \bibfield  {author} {\bibinfo {author} {\bibfnamefont {E.}~\bibnamefont
  {Kolomeitsev}}\ and\ \bibinfo {author} {\bibfnamefont {M.}~\bibnamefont
  {Lutz}},\ }\href {\doibase 10.1016/j.physletb.2003.10.118} {\bibfield
  {journal} {\bibinfo  {journal} {Phys. Lett. B}\ }\textbf {\bibinfo {volume}
  {582}},\ \bibinfo {pages} {39} (\bibinfo {year} {2004})},\ \Eprint
  {http://arxiv.org/abs/hep-ph/0307133} {arXiv:hep-ph/0307133} \BibitemShut
  {NoStop}%
\bibitem [{\citenamefont {Guo}\ \emph {et~al.}(2006)\citenamefont {Guo},
  \citenamefont {Shen}, \citenamefont {Chiang}, \citenamefont {Ping},\ and\
  \citenamefont {Zou}}]{Guo:2006fu}%
  \BibitemOpen
  \bibfield  {author} {\bibinfo {author} {\bibfnamefont {F.-K.}\ \bibnamefont
  {Guo}}, \bibinfo {author} {\bibfnamefont {P.-N.}\ \bibnamefont {Shen}},
  \bibinfo {author} {\bibfnamefont {H.-C.}\ \bibnamefont {Chiang}}, \bibinfo
  {author} {\bibfnamefont {R.-G.}\ \bibnamefont {Ping}}, \ and\ \bibinfo
  {author} {\bibfnamefont {B.-S.}\ \bibnamefont {Zou}},\ }\href {\doibase
  10.1016/j.physletb.2006.08.064} {\bibfield  {journal} {\bibinfo  {journal}
  {Phys. Lett.}\ }\textbf {\bibinfo {volume} {B641}},\ \bibinfo {pages} {278}
  (\bibinfo {year} {2006})},\ \Eprint {http://arxiv.org/abs/hep-ph/0603072}
  {arXiv:hep-ph/0603072 [hep-ph]} \BibitemShut {NoStop}%
\bibitem [{\citenamefont {Gamermann}\ \emph {et~al.}(2007)\citenamefont
  {Gamermann}, \citenamefont {Oset}, \citenamefont {Strottman},\ and\
  \citenamefont {Vicente~Vacas}}]{Gamermann:2006nm}%
  \BibitemOpen
  \bibfield  {author} {\bibinfo {author} {\bibfnamefont {D.}~\bibnamefont
  {Gamermann}}, \bibinfo {author} {\bibfnamefont {E.}~\bibnamefont {Oset}},
  \bibinfo {author} {\bibfnamefont {D.}~\bibnamefont {Strottman}}, \ and\
  \bibinfo {author} {\bibfnamefont {M.~J.}\ \bibnamefont {Vicente~Vacas}},\
  }\href {\doibase 10.1103/PhysRevD.76.074016} {\bibfield  {journal} {\bibinfo
  {journal} {Phys. Rev.}\ }\textbf {\bibinfo {volume} {D76}},\ \bibinfo {pages}
  {074016} (\bibinfo {year} {2007})},\ \Eprint
  {http://arxiv.org/abs/hep-ph/0612179} {arXiv:hep-ph/0612179 [hep-ph]}
  \BibitemShut {NoStop}%
\bibitem [{\citenamefont {Faessler}\ \emph
  {et~al.}(2007{\natexlab{a}})\citenamefont {Faessler}, \citenamefont
  {Gutsche}, \citenamefont {Lyubovitskij},\ and\ \citenamefont
  {Ma}}]{Faessler:2007gv}%
  \BibitemOpen
  \bibfield  {author} {\bibinfo {author} {\bibfnamefont {A.}~\bibnamefont
  {Faessler}}, \bibinfo {author} {\bibfnamefont {T.}~\bibnamefont {Gutsche}},
  \bibinfo {author} {\bibfnamefont {V.~E.}\ \bibnamefont {Lyubovitskij}}, \
  and\ \bibinfo {author} {\bibfnamefont {Y.-L.}\ \bibnamefont {Ma}},\ }\href
  {\doibase 10.1103/PhysRevD.76.014005} {\bibfield  {journal} {\bibinfo
  {journal} {Phys. Rev. D}\ }\textbf {\bibinfo {volume} {76}},\ \bibinfo
  {pages} {014005} (\bibinfo {year} {2007}{\natexlab{a}})},\ \Eprint
  {http://arxiv.org/abs/0705.0254} {arXiv:0705.0254 [hep-ph]} \BibitemShut
  {NoStop}%
\bibitem [{\citenamefont {Faessler}\ \emph
  {et~al.}(2007{\natexlab{b}})\citenamefont {Faessler}, \citenamefont
  {Gutsche}, \citenamefont {Lyubovitskij},\ and\ \citenamefont
  {Ma}}]{Faessler:2007us}%
  \BibitemOpen
  \bibfield  {author} {\bibinfo {author} {\bibfnamefont {A.}~\bibnamefont
  {Faessler}}, \bibinfo {author} {\bibfnamefont {T.}~\bibnamefont {Gutsche}},
  \bibinfo {author} {\bibfnamefont {V.~E.}\ \bibnamefont {Lyubovitskij}}, \
  and\ \bibinfo {author} {\bibfnamefont {Y.-L.}\ \bibnamefont {Ma}},\ }\href
  {\doibase 10.1103/PhysRevD.76.114008} {\bibfield  {journal} {\bibinfo
  {journal} {Phys. Rev. D}\ }\textbf {\bibinfo {volume} {76}},\ \bibinfo
  {pages} {114008} (\bibinfo {year} {2007}{\natexlab{b}})},\ \Eprint
  {http://arxiv.org/abs/0709.3946} {arXiv:0709.3946 [hep-ph]} \BibitemShut
  {NoStop}%
\bibitem [{\citenamefont {Liu}\ \emph {et~al.}(2013)\citenamefont {Liu},
  \citenamefont {Orginos}, \citenamefont {Guo}, \citenamefont {Hanhart},\ and\
  \citenamefont {Meissner}}]{Liu:2012zya}%
  \BibitemOpen
  \bibfield  {author} {\bibinfo {author} {\bibfnamefont {L.}~\bibnamefont
  {Liu}}, \bibinfo {author} {\bibfnamefont {K.}~\bibnamefont {Orginos}},
  \bibinfo {author} {\bibfnamefont {F.-K.}\ \bibnamefont {Guo}}, \bibinfo
  {author} {\bibfnamefont {C.}~\bibnamefont {Hanhart}}, \ and\ \bibinfo
  {author} {\bibfnamefont {U.-G.}\ \bibnamefont {Meissner}},\ }\href {\doibase
  10.1103/PhysRevD.87.014508} {\bibfield  {journal} {\bibinfo  {journal} {Phys.
  Rev.}\ }\textbf {\bibinfo {volume} {D87}},\ \bibinfo {pages} {014508}
  (\bibinfo {year} {2013})},\ \Eprint {http://arxiv.org/abs/1208.4535}
  {arXiv:1208.4535 [hep-lat]} \BibitemShut {NoStop}%
\bibitem [{\citenamefont {Altenbuchinger}\ \emph {et~al.}(2014)\citenamefont
  {Altenbuchinger}, \citenamefont {Geng},\ and\ \citenamefont
  {Weise}}]{Altenbuchinger:2013vwa}%
  \BibitemOpen
  \bibfield  {author} {\bibinfo {author} {\bibfnamefont {M.}~\bibnamefont
  {Altenbuchinger}}, \bibinfo {author} {\bibfnamefont {L.~S.}\ \bibnamefont
  {Geng}}, \ and\ \bibinfo {author} {\bibfnamefont {W.}~\bibnamefont {Weise}},\
  }\href {\doibase 10.1103/PhysRevD.89.014026} {\bibfield  {journal} {\bibinfo
  {journal} {Phys. Rev. D}\ }\textbf {\bibinfo {volume} {89}},\ \bibinfo
  {pages} {014026} (\bibinfo {year} {2014})},\ \Eprint
  {http://arxiv.org/abs/1309.4743} {arXiv:1309.4743 [hep-ph]} \BibitemShut
  {NoStop}%
\bibitem [{\citenamefont {Dong}\ \emph {et~al.}(2013)\citenamefont {Dong},
  \citenamefont {Faessler}, \citenamefont {Gutsche},\ and\ \citenamefont
  {Lyubovitskij}}]{Dong:2013iqa}%
  \BibitemOpen
  \bibfield  {author} {\bibinfo {author} {\bibfnamefont {Y.}~\bibnamefont
  {Dong}}, \bibinfo {author} {\bibfnamefont {A.}~\bibnamefont {Faessler}},
  \bibinfo {author} {\bibfnamefont {T.}~\bibnamefont {Gutsche}}, \ and\
  \bibinfo {author} {\bibfnamefont {V.~E.}\ \bibnamefont {Lyubovitskij}},\
  }\href {\doibase 10.1103/PhysRevD.88.014030} {\bibfield  {journal} {\bibinfo
  {journal} {Phys. Rev. D}\ }\textbf {\bibinfo {volume} {88}},\ \bibinfo
  {pages} {014030} (\bibinfo {year} {2013})},\ \Eprint
  {http://arxiv.org/abs/1306.0824} {arXiv:1306.0824 [hep-ph]} \BibitemShut
  {NoStop}%
\bibitem [{\citenamefont {Wang}\ \emph
  {et~al.}(2020{\natexlab{a}})\citenamefont {Wang}, \citenamefont {Meng},\ and\
  \citenamefont {Zhu}}]{Wang:2020dko}%
  \BibitemOpen
  \bibfield  {author} {\bibinfo {author} {\bibfnamefont {B.}~\bibnamefont
  {Wang}}, \bibinfo {author} {\bibfnamefont {L.}~\bibnamefont {Meng}}, \ and\
  \bibinfo {author} {\bibfnamefont {S.-L.}\ \bibnamefont {Zhu}},\ }\href@noop
  {} {\  (\bibinfo {year} {2020}{\natexlab{a}})},\ \Eprint
  {http://arxiv.org/abs/2009.01980} {arXiv:2009.01980 [hep-ph]} \BibitemShut
  {NoStop}%
\bibitem [{\citenamefont {Yang}\ \emph
  {et~al.}(2020{\natexlab{a}})\citenamefont {Yang}, \citenamefont {Cao},
  \citenamefont {Guo}, \citenamefont {Nieves},\ and\ \citenamefont
  {Valderrama}}]{Yang:2020nrt}%
  \BibitemOpen
  \bibfield  {author} {\bibinfo {author} {\bibfnamefont {Z.}~\bibnamefont
  {Yang}}, \bibinfo {author} {\bibfnamefont {X.}~\bibnamefont {Cao}}, \bibinfo
  {author} {\bibfnamefont {F.-K.}\ \bibnamefont {Guo}}, \bibinfo {author}
  {\bibfnamefont {J.}~\bibnamefont {Nieves}}, \ and\ \bibinfo {author}
  {\bibfnamefont {M.~P.}\ \bibnamefont {Valderrama}},\ }\href@noop {} {\
  (\bibinfo {year} {2020}{\natexlab{a}})},\ \Eprint
  {http://arxiv.org/abs/2011.08725} {arXiv:2011.08725 [hep-ph]} \BibitemShut
  {NoStop}%
\bibitem [{\citenamefont {Meng}\ \emph {et~al.}(2020)\citenamefont {Meng},
  \citenamefont {Wang},\ and\ \citenamefont {Zhu}}]{Meng:2020ihj}%
  \BibitemOpen
  \bibfield  {author} {\bibinfo {author} {\bibfnamefont {L.}~\bibnamefont
  {Meng}}, \bibinfo {author} {\bibfnamefont {B.}~\bibnamefont {Wang}}, \ and\
  \bibinfo {author} {\bibfnamefont {S.-L.}\ \bibnamefont {Zhu}},\ }\href@noop
  {} {\  (\bibinfo {year} {2020})},\ \Eprint {http://arxiv.org/abs/2011.08656}
  {arXiv:2011.08656 [hep-ph]} \BibitemShut {NoStop}%
\bibitem [{\citenamefont {Du}\ \emph {et~al.}(2020{\natexlab{a}})\citenamefont
  {Du}, \citenamefont {Wang},\ and\ \citenamefont {Zhao}}]{Du:2020vwb}%
  \BibitemOpen
  \bibfield  {author} {\bibinfo {author} {\bibfnamefont {M.-C.}\ \bibnamefont
  {Du}}, \bibinfo {author} {\bibfnamefont {Q.}~\bibnamefont {Wang}}, \ and\
  \bibinfo {author} {\bibfnamefont {Q.}~\bibnamefont {Zhao}},\ }\href@noop {}
  {\  (\bibinfo {year} {2020}{\natexlab{a}})},\ \Eprint
  {http://arxiv.org/abs/2011.09225} {arXiv:2011.09225 [hep-ph]} \BibitemShut
  {NoStop}%
\bibitem [{\citenamefont {Liu}\ \emph {et~al.}(2019{\natexlab{b}})\citenamefont
  {Liu}, \citenamefont {Pan}, \citenamefont {Peng}, \citenamefont
  {Sánchez~Sánchez}, \citenamefont {Geng}, \citenamefont {Hosaka},\ and\
  \citenamefont {Pavon~Valderrama}}]{Liu:2019tjn}%
  \BibitemOpen
  \bibfield  {author} {\bibinfo {author} {\bibfnamefont {M.-Z.}\ \bibnamefont
  {Liu}}, \bibinfo {author} {\bibfnamefont {Y.-W.}\ \bibnamefont {Pan}},
  \bibinfo {author} {\bibfnamefont {F.-Z.}\ \bibnamefont {Peng}}, \bibinfo
  {author} {\bibfnamefont {M.}~\bibnamefont {Sánchez~Sánchez}}, \bibinfo
  {author} {\bibfnamefont {L.-S.}\ \bibnamefont {Geng}}, \bibinfo {author}
  {\bibfnamefont {A.}~\bibnamefont {Hosaka}}, \ and\ \bibinfo {author}
  {\bibfnamefont {M.}~\bibnamefont {Pavon~Valderrama}},\ }\href {\doibase
  10.1103/PhysRevLett.122.242001} {\bibfield  {journal} {\bibinfo  {journal}
  {Phys. Rev. Lett.}\ }\textbf {\bibinfo {volume} {122}},\ \bibinfo {pages}
  {242001} (\bibinfo {year} {2019}{\natexlab{b}})},\ \Eprint
  {http://arxiv.org/abs/1903.11560} {arXiv:1903.11560 [hep-ph]} \BibitemShut
  {NoStop}%
\bibitem [{\citenamefont {Du}\ \emph {et~al.}(2020{\natexlab{b}})\citenamefont
  {Du}, \citenamefont {Baru}, \citenamefont {Guo}, \citenamefont {Hanhart},
  \citenamefont {Meißner}, \citenamefont {Oller},\ and\ \citenamefont
  {Wang}}]{Du:2019pij}%
  \BibitemOpen
  \bibfield  {author} {\bibinfo {author} {\bibfnamefont {M.-L.}\ \bibnamefont
  {Du}}, \bibinfo {author} {\bibfnamefont {V.}~\bibnamefont {Baru}}, \bibinfo
  {author} {\bibfnamefont {F.-K.}\ \bibnamefont {Guo}}, \bibinfo {author}
  {\bibfnamefont {C.}~\bibnamefont {Hanhart}}, \bibinfo {author} {\bibfnamefont
  {U.-G.}\ \bibnamefont {Meißner}}, \bibinfo {author} {\bibfnamefont {J.~A.}\
  \bibnamefont {Oller}}, \ and\ \bibinfo {author} {\bibfnamefont
  {Q.}~\bibnamefont {Wang}},\ }\href {\doibase 10.1103/PhysRevLett.124.072001}
  {\bibfield  {journal} {\bibinfo  {journal} {Phys. Rev. Lett.}\ }\textbf
  {\bibinfo {volume} {124}},\ \bibinfo {pages} {072001} (\bibinfo {year}
  {2020}{\natexlab{b}})},\ \Eprint {http://arxiv.org/abs/1910.11846}
  {arXiv:1910.11846 [hep-ph]} \BibitemShut {NoStop}%
\bibitem [{\citenamefont {Xiao}\ \emph
  {et~al.}(2019{\natexlab{a}})\citenamefont {Xiao}, \citenamefont {Nieves},\
  and\ \citenamefont {Oset}}]{Xiao:2019aya}%
  \BibitemOpen
  \bibfield  {author} {\bibinfo {author} {\bibfnamefont {C.~W.}\ \bibnamefont
  {Xiao}}, \bibinfo {author} {\bibfnamefont {J.}~\bibnamefont {Nieves}}, \ and\
  \bibinfo {author} {\bibfnamefont {E.}~\bibnamefont {Oset}},\ }\href {\doibase
  10.1103/PhysRevD.100.014021} {\bibfield  {journal} {\bibinfo  {journal}
  {Phys. Rev.}\ }\textbf {\bibinfo {volume} {D100}},\ \bibinfo {pages} {014021}
  (\bibinfo {year} {2019}{\natexlab{a}})},\ \Eprint
  {http://arxiv.org/abs/1904.01296} {arXiv:1904.01296 [hep-ph]} \BibitemShut
  {NoStop}%
\bibitem [{\citenamefont {Sakai}\ \emph {et~al.}(2019)\citenamefont {Sakai},
  \citenamefont {Jing},\ and\ \citenamefont {Guo}}]{Sakai:2019qph}%
  \BibitemOpen
  \bibfield  {author} {\bibinfo {author} {\bibfnamefont {S.}~\bibnamefont
  {Sakai}}, \bibinfo {author} {\bibfnamefont {H.-J.}\ \bibnamefont {Jing}}, \
  and\ \bibinfo {author} {\bibfnamefont {F.-K.}\ \bibnamefont {Guo}},\ }\href
  {\doibase 10.1103/PhysRevD.100.074007} {\bibfield  {journal} {\bibinfo
  {journal} {Phys. Rev.}\ }\textbf {\bibinfo {volume} {D100}},\ \bibinfo
  {pages} {074007} (\bibinfo {year} {2019})},\ \Eprint
  {http://arxiv.org/abs/1907.03414} {arXiv:1907.03414 [hep-ph]} \BibitemShut
  {NoStop}%
\bibitem [{\citenamefont {Yamaguchi}\ \emph {et~al.}(2019)\citenamefont
  {Yamaguchi}, \citenamefont {García-Tecocoatzi}, \citenamefont {Giachino},
  \citenamefont {Hosaka}, \citenamefont {Santopinto}, \citenamefont
  {Takeuchi},\ and\ \citenamefont {Takizawa}}]{Yamaguchi:2019seo}%
  \BibitemOpen
  \bibfield  {author} {\bibinfo {author} {\bibfnamefont {Y.}~\bibnamefont
  {Yamaguchi}}, \bibinfo {author} {\bibfnamefont {H.}~\bibnamefont
  {García-Tecocoatzi}}, \bibinfo {author} {\bibfnamefont {A.}~\bibnamefont
  {Giachino}}, \bibinfo {author} {\bibfnamefont {A.}~\bibnamefont {Hosaka}},
  \bibinfo {author} {\bibfnamefont {E.}~\bibnamefont {Santopinto}}, \bibinfo
  {author} {\bibfnamefont {S.}~\bibnamefont {Takeuchi}}, \ and\ \bibinfo
  {author} {\bibfnamefont {M.}~\bibnamefont {Takizawa}},\ }\href@noop {} {\
  (\bibinfo {year} {2019})},\ \Eprint {http://arxiv.org/abs/1907.04684}
  {arXiv:1907.04684 [hep-ph]} \BibitemShut {NoStop}%
\bibitem [{\citenamefont {Lin}\ and\ \citenamefont {Zou}(2019)}]{Lin:2019qiv}%
  \BibitemOpen
  \bibfield  {author} {\bibinfo {author} {\bibfnamefont {Y.-H.}\ \bibnamefont
  {Lin}}\ and\ \bibinfo {author} {\bibfnamefont {B.-S.}\ \bibnamefont {Zou}},\
  }\href {\doibase 10.1103/PhysRevD.100.056005} {\bibfield  {journal} {\bibinfo
   {journal} {Phys. Rev.}\ }\textbf {\bibinfo {volume} {D100}},\ \bibinfo
  {pages} {056005} (\bibinfo {year} {2019})},\ \Eprint
  {http://arxiv.org/abs/1908.05309} {arXiv:1908.05309 [hep-ph]} \BibitemShut
  {NoStop}%
\bibitem [{\citenamefont {Pavon~Valderrama}(2019)}]{Valderrama:2019chc}%
  \BibitemOpen
  \bibfield  {author} {\bibinfo {author} {\bibfnamefont {M.}~\bibnamefont
  {Pavon~Valderrama}},\ }\href {\doibase 10.1103/PhysRevD.100.094028}
  {\bibfield  {journal} {\bibinfo  {journal} {Phys. Rev.}\ }\textbf {\bibinfo
  {volume} {D100}},\ \bibinfo {pages} {094028} (\bibinfo {year} {2019})},\
  \Eprint {http://arxiv.org/abs/1907.05294} {arXiv:1907.05294 [hep-ph]}
  \BibitemShut {NoStop}%
\bibitem [{\citenamefont {Liu}\ \emph {et~al.}(2019{\natexlab{c}})\citenamefont
  {Liu}, \citenamefont {Wu}, \citenamefont {Sánchez~Sánchez}, \citenamefont
  {Valderrama}, \citenamefont {Geng},\ and\ \citenamefont {Xie}}]{Liu:2019zvb}%
  \BibitemOpen
  \bibfield  {author} {\bibinfo {author} {\bibfnamefont {M.-Z.}\ \bibnamefont
  {Liu}}, \bibinfo {author} {\bibfnamefont {T.-W.}\ \bibnamefont {Wu}},
  \bibinfo {author} {\bibfnamefont {M.}~\bibnamefont {Sánchez~Sánchez}},
  \bibinfo {author} {\bibfnamefont {M.~P.}\ \bibnamefont {Valderrama}},
  \bibinfo {author} {\bibfnamefont {L.-S.}\ \bibnamefont {Geng}}, \ and\
  \bibinfo {author} {\bibfnamefont {J.-J.}\ \bibnamefont {Xie}},\ }\href@noop
  {} {\  (\bibinfo {year} {2019}{\natexlab{c}})},\ \Eprint
  {http://arxiv.org/abs/1907.06093} {arXiv:1907.06093 [hep-ph]} \BibitemShut
  {NoStop}%
\bibitem [{\citenamefont {Savage}\ and\ \citenamefont
  {Wise}(1990)}]{Savage:1990di}%
  \BibitemOpen
  \bibfield  {author} {\bibinfo {author} {\bibfnamefont {M.~J.}\ \bibnamefont
  {Savage}}\ and\ \bibinfo {author} {\bibfnamefont {M.~B.}\ \bibnamefont
  {Wise}},\ }\href {\doibase 10.1016/0370-2693(90)90035-5} {\bibfield
  {journal} {\bibinfo  {journal} {Phys. Lett. B}\ }\textbf {\bibinfo {volume}
  {248}},\ \bibinfo {pages} {177} (\bibinfo {year} {1990})}\BibitemShut
  {NoStop}%
\bibitem [{\citenamefont {Hu}\ and\ \citenamefont {Mehen}(2006)}]{Hu:2005gf}%
  \BibitemOpen
  \bibfield  {author} {\bibinfo {author} {\bibfnamefont {J.}~\bibnamefont
  {Hu}}\ and\ \bibinfo {author} {\bibfnamefont {T.}~\bibnamefont {Mehen}},\
  }\href {\doibase 10.1103/PhysRevD.73.054003} {\bibfield  {journal} {\bibinfo
  {journal} {Phys. Rev. D}\ }\textbf {\bibinfo {volume} {73}},\ \bibinfo
  {pages} {054003} (\bibinfo {year} {2006})},\ \Eprint
  {http://arxiv.org/abs/hep-ph/0511321} {arXiv:hep-ph/0511321} \BibitemShut
  {NoStop}%
\bibitem [{\citenamefont {Padmanath}\ \emph {et~al.}(2015)\citenamefont
  {Padmanath}, \citenamefont {Edwards}, \citenamefont {Mathur},\ and\
  \citenamefont {Peardon}}]{Padmanath:2015jea}%
  \BibitemOpen
  \bibfield  {author} {\bibinfo {author} {\bibfnamefont {M.}~\bibnamefont
  {Padmanath}}, \bibinfo {author} {\bibfnamefont {R.~G.}\ \bibnamefont
  {Edwards}}, \bibinfo {author} {\bibfnamefont {N.}~\bibnamefont {Mathur}}, \
  and\ \bibinfo {author} {\bibfnamefont {M.}~\bibnamefont {Peardon}},\ }\href
  {\doibase 10.1103/PhysRevD.91.094502} {\bibfield  {journal} {\bibinfo
  {journal} {Phys. Rev. D}\ }\textbf {\bibinfo {volume} {91}},\ \bibinfo
  {pages} {094502} (\bibinfo {year} {2015})},\ \Eprint
  {http://arxiv.org/abs/1502.01845} {arXiv:1502.01845 [hep-lat]} \BibitemShut
  {NoStop}%
\bibitem [{\citenamefont {Mathur}\ and\ \citenamefont
  {Padmanath}(2019)}]{Mathur:2018rwu}%
  \BibitemOpen
  \bibfield  {author} {\bibinfo {author} {\bibfnamefont {N.}~\bibnamefont
  {Mathur}}\ and\ \bibinfo {author} {\bibfnamefont {M.}~\bibnamefont
  {Padmanath}},\ }\href {\doibase 10.1103/PhysRevD.99.031501} {\bibfield
  {journal} {\bibinfo  {journal} {Phys. Rev. D}\ }\textbf {\bibinfo {volume}
  {99}},\ \bibinfo {pages} {031501} (\bibinfo {year} {2019})},\ \Eprint
  {http://arxiv.org/abs/1807.00174} {arXiv:1807.00174 [hep-lat]} \BibitemShut
  {NoStop}%
\bibitem [{\citenamefont {Roncaglia}\ \emph {et~al.}(1995)\citenamefont
  {Roncaglia}, \citenamefont {Lichtenberg},\ and\ \citenamefont
  {Predazzi}}]{Roncaglia:1995az}%
  \BibitemOpen
  \bibfield  {author} {\bibinfo {author} {\bibfnamefont {R.}~\bibnamefont
  {Roncaglia}}, \bibinfo {author} {\bibfnamefont {D.}~\bibnamefont
  {Lichtenberg}}, \ and\ \bibinfo {author} {\bibfnamefont {E.}~\bibnamefont
  {Predazzi}},\ }\href {\doibase 10.1103/PhysRevD.52.1722} {\bibfield
  {journal} {\bibinfo  {journal} {Phys. Rev. D}\ }\textbf {\bibinfo {volume}
  {52}},\ \bibinfo {pages} {1722} (\bibinfo {year} {1995})},\ \Eprint
  {http://arxiv.org/abs/hep-ph/9502251} {arXiv:hep-ph/9502251} \BibitemShut
  {NoStop}%
\bibitem [{\citenamefont {Karliner}\ and\ \citenamefont
  {Rosner}(2014)}]{Karliner:2014gca}%
  \BibitemOpen
  \bibfield  {author} {\bibinfo {author} {\bibfnamefont {M.}~\bibnamefont
  {Karliner}}\ and\ \bibinfo {author} {\bibfnamefont {J.~L.}\ \bibnamefont
  {Rosner}},\ }\href {\doibase 10.1103/PhysRevD.90.094007} {\bibfield
  {journal} {\bibinfo  {journal} {Phys. Rev. D}\ }\textbf {\bibinfo {volume}
  {90}},\ \bibinfo {pages} {094007} (\bibinfo {year} {2014})},\ \Eprint
  {http://arxiv.org/abs/1408.5877} {arXiv:1408.5877 [hep-ph]} \BibitemShut
  {NoStop}%
\bibitem [{\citenamefont {Pan}\ \emph {et~al.}(2019)\citenamefont {Pan},
  \citenamefont {Liu}, \citenamefont {Peng}, \citenamefont {Sánchez~Sánchez},
  \citenamefont {Geng},\ and\ \citenamefont {Pavon~Valderrama}}]{Pan:2019skd}%
  \BibitemOpen
  \bibfield  {author} {\bibinfo {author} {\bibfnamefont {Y.-W.}\ \bibnamefont
  {Pan}}, \bibinfo {author} {\bibfnamefont {M.-Z.}\ \bibnamefont {Liu}},
  \bibinfo {author} {\bibfnamefont {F.-Z.}\ \bibnamefont {Peng}}, \bibinfo
  {author} {\bibfnamefont {M.}~\bibnamefont {Sánchez~Sánchez}}, \bibinfo
  {author} {\bibfnamefont {L.-S.}\ \bibnamefont {Geng}}, \ and\ \bibinfo
  {author} {\bibfnamefont {M.}~\bibnamefont {Pavon~Valderrama}},\ }\href@noop
  {} {\  (\bibinfo {year} {2019})},\ \Eprint {http://arxiv.org/abs/1907.11220}
  {arXiv:1907.11220 [hep-ph]} \BibitemShut {NoStop}%
\bibitem [{\citenamefont {Pan}\ \emph {et~al.}(2020)\citenamefont {Pan},
  \citenamefont {Liu},\ and\ \citenamefont {Geng}}]{Pan:2020xek}%
  \BibitemOpen
  \bibfield  {author} {\bibinfo {author} {\bibfnamefont {Y.-W.}\ \bibnamefont
  {Pan}}, \bibinfo {author} {\bibfnamefont {M.-Z.}\ \bibnamefont {Liu}}, \ and\
  \bibinfo {author} {\bibfnamefont {L.-S.}\ \bibnamefont {Geng}},\ }\href
  {\doibase 10.1103/PhysRevD.102.054025} {\bibfield  {journal} {\bibinfo
  {journal} {Phys. Rev. D}\ }\textbf {\bibinfo {volume} {102}},\ \bibinfo
  {pages} {054025} (\bibinfo {year} {2020})},\ \Eprint
  {http://arxiv.org/abs/2004.07467} {arXiv:2004.07467 [hep-ph]} \BibitemShut
  {NoStop}%
\bibitem [{\citenamefont {Guo}\ \emph {et~al.}(2013{\natexlab{b}})\citenamefont
  {Guo}, \citenamefont {Hidalgo-Duque}, \citenamefont {Nieves},\ and\
  \citenamefont {Valderrama}}]{Guo:2013xga}%
  \BibitemOpen
  \bibfield  {author} {\bibinfo {author} {\bibfnamefont {F.-K.}\ \bibnamefont
  {Guo}}, \bibinfo {author} {\bibfnamefont {C.}~\bibnamefont {Hidalgo-Duque}},
  \bibinfo {author} {\bibfnamefont {J.}~\bibnamefont {Nieves}}, \ and\ \bibinfo
  {author} {\bibfnamefont {M.~P.}\ \bibnamefont {Valderrama}},\ }\href
  {\doibase 10.1103/PhysRevD.88.054014} {\bibfield  {journal} {\bibinfo
  {journal} {Phys. Rev.}\ }\textbf {\bibinfo {volume} {D88}},\ \bibinfo {pages}
  {054014} (\bibinfo {year} {2013}{\natexlab{b}})},\ \Eprint
  {http://arxiv.org/abs/1305.4052} {arXiv:1305.4052 [hep-ph]} \BibitemShut
  {NoStop}%
\bibitem [{\citenamefont {Peng}\ \emph
  {et~al.}(2020{\natexlab{a}})\citenamefont {Peng}, \citenamefont {Liu},
  \citenamefont {S\'anchez~S\'anchez},\ and\ \citenamefont
  {Pavon~Valderrama}}]{Peng:2020xrf}%
  \BibitemOpen
  \bibfield  {author} {\bibinfo {author} {\bibfnamefont {F.-Z.}\ \bibnamefont
  {Peng}}, \bibinfo {author} {\bibfnamefont {M.-Z.}\ \bibnamefont {Liu}},
  \bibinfo {author} {\bibfnamefont {M.}~\bibnamefont {S\'anchez~S\'anchez}}, \
  and\ \bibinfo {author} {\bibfnamefont {M.}~\bibnamefont {Pavon~Valderrama}},\
  }\href@noop {} {\  (\bibinfo {year} {2020}{\natexlab{a}})},\ \Eprint
  {http://arxiv.org/abs/2004.05658} {arXiv:2004.05658 [hep-ph]} \BibitemShut
  {NoStop}%
\bibitem [{\citenamefont {Peng}\ \emph
  {et~al.}(2020{\natexlab{b}})\citenamefont {Peng}, \citenamefont {Yan},
  \citenamefont {S\'anchez~S\'anchez},\ and\ \citenamefont
  {Valderrama}}]{Peng:2020hql}%
  \BibitemOpen
  \bibfield  {author} {\bibinfo {author} {\bibfnamefont {F.-Z.}\ \bibnamefont
  {Peng}}, \bibinfo {author} {\bibfnamefont {M.-J.}\ \bibnamefont {Yan}},
  \bibinfo {author} {\bibfnamefont {M.}~\bibnamefont {S\'anchez~S\'anchez}}, \
  and\ \bibinfo {author} {\bibfnamefont {M.~P.}\ \bibnamefont {Valderrama}},\
  }\href@noop {} {\  (\bibinfo {year} {2020}{\natexlab{b}})},\ \Eprint
  {http://arxiv.org/abs/2011.01915} {arXiv:2011.01915 [hep-ph]} \BibitemShut
  {NoStop}%
\bibitem [{\citenamefont {Xiao}\ \emph
  {et~al.}(2019{\natexlab{b}})\citenamefont {Xiao}, \citenamefont {Nieves},\
  and\ \citenamefont {Oset}}]{Xiao:2019gjd}%
  \BibitemOpen
  \bibfield  {author} {\bibinfo {author} {\bibfnamefont {C.}~\bibnamefont
  {Xiao}}, \bibinfo {author} {\bibfnamefont {J.}~\bibnamefont {Nieves}}, \ and\
  \bibinfo {author} {\bibfnamefont {E.}~\bibnamefont {Oset}},\ }\href {\doibase
  10.1016/j.physletb.2019.135051} {\bibfield  {journal} {\bibinfo  {journal}
  {Phys. Lett. B}\ }\textbf {\bibinfo {volume} {799}},\ \bibinfo {pages}
  {135051} (\bibinfo {year} {2019}{\natexlab{b}})},\ \Eprint
  {http://arxiv.org/abs/1906.09010} {arXiv:1906.09010 [hep-ph]} \BibitemShut
  {NoStop}%
\bibitem [{\citenamefont {Liu}\ \emph {et~al.}(2020)\citenamefont {Liu},
  \citenamefont {Pan},\ and\ \citenamefont {Geng}}]{Liu:2020hcv}%
  \BibitemOpen
  \bibfield  {author} {\bibinfo {author} {\bibfnamefont {M.-Z.}\ \bibnamefont
  {Liu}}, \bibinfo {author} {\bibfnamefont {Y.-W.}\ \bibnamefont {Pan}}, \ and\
  \bibinfo {author} {\bibfnamefont {L.-S.}\ \bibnamefont {Geng}},\ }\href@noop
  {} {\  (\bibinfo {year} {2020})},\ \Eprint {http://arxiv.org/abs/2011.07935}
  {arXiv:2011.07935 [hep-ph]} \BibitemShut {NoStop}%
\bibitem [{\citenamefont {Aaij}\ \emph {et~al.}(2020)\citenamefont {Aaij} \emph
  {et~al.}}]{Aaij:2020fnh}%
  \BibitemOpen
  \bibfield  {author} {\bibinfo {author} {\bibfnamefont {R.}~\bibnamefont
  {Aaij}} \emph {et~al.} (\bibinfo {collaboration} {LHCb}),\ }\href {\doibase
  10.1016/j.scib.2020.08.032} {\bibfield  {journal} {\bibinfo  {journal} {Sci.
  Bull.}\ }\textbf {\bibinfo {volume} {2020}},\ \bibinfo {pages} {65} (\bibinfo
  {year} {2020})},\ \Eprint {http://arxiv.org/abs/2006.16957} {arXiv:2006.16957
  [hep-ex]} \BibitemShut {NoStop}%
\bibitem [{\citenamefont {Ma}\ and\ \citenamefont {Zhang}(2020)}]{Ma:2020kwb}%
  \BibitemOpen
  \bibfield  {author} {\bibinfo {author} {\bibfnamefont {Y.-Q.}\ \bibnamefont
  {Ma}}\ and\ \bibinfo {author} {\bibfnamefont {H.-F.}\ \bibnamefont {Zhang}},\
  }\href@noop {} {\  (\bibinfo {year} {2020})},\ \Eprint
  {http://arxiv.org/abs/2009.08376} {arXiv:2009.08376 [hep-ph]} \BibitemShut
  {NoStop}%
\bibitem [{\citenamefont {Dong}\ \emph {et~al.}(2020)\citenamefont {Dong},
  \citenamefont {Baru}, \citenamefont {Guo}, \citenamefont {Hanhart},\ and\
  \citenamefont {Nefediev}}]{Dong:2020nwy}%
  \BibitemOpen
  \bibfield  {author} {\bibinfo {author} {\bibfnamefont {X.-K.}\ \bibnamefont
  {Dong}}, \bibinfo {author} {\bibfnamefont {V.}~\bibnamefont {Baru}}, \bibinfo
  {author} {\bibfnamefont {F.-K.}\ \bibnamefont {Guo}}, \bibinfo {author}
  {\bibfnamefont {C.}~\bibnamefont {Hanhart}}, \ and\ \bibinfo {author}
  {\bibfnamefont {A.}~\bibnamefont {Nefediev}},\ }\href@noop {} {\  (\bibinfo
  {year} {2020})},\ \Eprint {http://arxiv.org/abs/2009.07795} {arXiv:2009.07795
  [hep-ph]} \BibitemShut {NoStop}%
\bibitem [{\citenamefont {Wang}(2020)}]{Wang:2020dlo}%
  \BibitemOpen
  \bibfield  {author} {\bibinfo {author} {\bibfnamefont {Z.-G.}\ \bibnamefont
  {Wang}},\ }\href@noop {} {\  (\bibinfo {year} {2020})},\ \Eprint
  {http://arxiv.org/abs/2009.05371} {arXiv:2009.05371 [hep-ph]} \BibitemShut
  {NoStop}%
\bibitem [{\citenamefont {Karliner}\ and\ \citenamefont
  {Rosner}(2020)}]{Karliner:2020dta}%
  \BibitemOpen
  \bibfield  {author} {\bibinfo {author} {\bibfnamefont {M.}~\bibnamefont
  {Karliner}}\ and\ \bibinfo {author} {\bibfnamefont {J.~L.}\ \bibnamefont
  {Rosner}},\ }\href@noop {} {\  (\bibinfo {year} {2020})},\ \Eprint
  {http://arxiv.org/abs/2009.04429} {arXiv:2009.04429 [hep-ph]} \BibitemShut
  {NoStop}%
\bibitem [{\citenamefont {Maciu\l{}a}\ \emph {et~al.}(2020)\citenamefont
  {Maciu\l{}a}, \citenamefont {Sch\"afer},\ and\ \citenamefont
  {Szczurek}}]{Maciula:2020wri}%
  \BibitemOpen
  \bibfield  {author} {\bibinfo {author} {\bibfnamefont {R.}~\bibnamefont
  {Maciu\l{}a}}, \bibinfo {author} {\bibfnamefont {W.}~\bibnamefont
  {Sch\"afer}}, \ and\ \bibinfo {author} {\bibfnamefont {A.}~\bibnamefont
  {Szczurek}},\ }\href@noop {} {\  (\bibinfo {year} {2020})},\ \Eprint
  {http://arxiv.org/abs/2009.02100} {arXiv:2009.02100 [hep-ph]} \BibitemShut
  {NoStop}%
\bibitem [{\citenamefont {Chao}\ and\ \citenamefont
  {Zhu}(2020)}]{Chao:2020dml}%
  \BibitemOpen
  \bibfield  {author} {\bibinfo {author} {\bibfnamefont {K.-T.}\ \bibnamefont
  {Chao}}\ and\ \bibinfo {author} {\bibfnamefont {S.-L.}\ \bibnamefont {Zhu}}\
  }(\bibinfo {year} {2020})\ \Eprint {http://arxiv.org/abs/2008.07670}
  {arXiv:2008.07670 [hep-ph]} \BibitemShut {NoStop}%
\bibitem [{\citenamefont {Richard}(2020)}]{Richard:2020hdw}%
  \BibitemOpen
  \bibfield  {author} {\bibinfo {author} {\bibfnamefont {J.-M.}\ \bibnamefont
  {Richard}},\ }\href@noop {} {\  (\bibinfo {year} {2020})},\ \Eprint
  {http://arxiv.org/abs/2008.01962} {arXiv:2008.01962 [hep-ph]} \BibitemShut
  {NoStop}%
\bibitem [{\citenamefont {Maiani}(2020)}]{Maiani:2020pur}%
  \BibitemOpen
  \bibfield  {author} {\bibinfo {author} {\bibfnamefont {L.}~\bibnamefont
  {Maiani}},\ }\href@noop {} {\  (\bibinfo {year} {2020})},\ \Eprint
  {http://arxiv.org/abs/2008.01637} {arXiv:2008.01637 [hep-ph]} \BibitemShut
  {NoStop}%
\bibitem [{\citenamefont {Sonnenschein}\ and\ \citenamefont
  {Weissman}(2020)}]{Sonnenschein:2020nwn}%
  \BibitemOpen
  \bibfield  {author} {\bibinfo {author} {\bibfnamefont {J.}~\bibnamefont
  {Sonnenschein}}\ and\ \bibinfo {author} {\bibfnamefont {D.}~\bibnamefont
  {Weissman}},\ }\href@noop {} {\  (\bibinfo {year} {2020})},\ \Eprint
  {http://arxiv.org/abs/2008.01095} {arXiv:2008.01095 [hep-ph]} \BibitemShut
  {NoStop}%
\bibitem [{\citenamefont {Giron}\ and\ \citenamefont
  {Lebed}(2020)}]{Giron:2020wpx}%
  \BibitemOpen
  \bibfield  {author} {\bibinfo {author} {\bibfnamefont {J.~F.}\ \bibnamefont
  {Giron}}\ and\ \bibinfo {author} {\bibfnamefont {R.~F.}\ \bibnamefont
  {Lebed}},\ }\href {\doibase 10.1103/PhysRevD.102.074003} {\bibfield
  {journal} {\bibinfo  {journal} {Phys. Rev. D}\ }\textbf {\bibinfo {volume}
  {102}},\ \bibinfo {pages} {074003} (\bibinfo {year} {2020})},\ \Eprint
  {http://arxiv.org/abs/2008.01631} {arXiv:2008.01631 [hep-ph]} \BibitemShut
  {NoStop}%
\bibitem [{\citenamefont {Wang}\ \emph
  {et~al.}(2020{\natexlab{b}})\citenamefont {Wang}, \citenamefont {Lin},
  \citenamefont {Xu}, \citenamefont {Xie}, \citenamefont {Huang},\ and\
  \citenamefont {Chen}}]{Wang:2020gmd}%
  \BibitemOpen
  \bibfield  {author} {\bibinfo {author} {\bibfnamefont {X.-Y.}\ \bibnamefont
  {Wang}}, \bibinfo {author} {\bibfnamefont {Q.-Y.}\ \bibnamefont {Lin}},
  \bibinfo {author} {\bibfnamefont {H.}~\bibnamefont {Xu}}, \bibinfo {author}
  {\bibfnamefont {Y.-P.}\ \bibnamefont {Xie}}, \bibinfo {author} {\bibfnamefont
  {Y.}~\bibnamefont {Huang}}, \ and\ \bibinfo {author} {\bibfnamefont
  {X.}~\bibnamefont {Chen}},\ }\href@noop {} {\  (\bibinfo {year}
  {2020}{\natexlab{b}})},\ \Eprint {http://arxiv.org/abs/2007.09697}
  {arXiv:2007.09697 [hep-ph]} \BibitemShut {NoStop}%
\bibitem [{\citenamefont {Weng}\ \emph {et~al.}(2020)\citenamefont {Weng},
  \citenamefont {Chen}, \citenamefont {Deng},\ and\ \citenamefont
  {Zhu}}]{Weng:2020jao}%
  \BibitemOpen
  \bibfield  {author} {\bibinfo {author} {\bibfnamefont {X.-Z.}\ \bibnamefont
  {Weng}}, \bibinfo {author} {\bibfnamefont {X.-L.}\ \bibnamefont {Chen}},
  \bibinfo {author} {\bibfnamefont {W.-Z.}\ \bibnamefont {Deng}}, \ and\
  \bibinfo {author} {\bibfnamefont {S.-L.}\ \bibnamefont {Zhu}},\ }\href@noop
  {} {\  (\bibinfo {year} {2020})},\ \Eprint {http://arxiv.org/abs/2010.05163}
  {arXiv:2010.05163 [hep-ph]} \BibitemShut {NoStop}%
\bibitem [{\citenamefont {Zhu}(2020)}]{Zhu:2020xni}%
  \BibitemOpen
  \bibfield  {author} {\bibinfo {author} {\bibfnamefont {R.}~\bibnamefont
  {Zhu}},\ }\href@noop {} {\  (\bibinfo {year} {2020})},\ \Eprint
  {http://arxiv.org/abs/2010.09082} {arXiv:2010.09082 [hep-ph]} \BibitemShut
  {NoStop}%
\bibitem [{\citenamefont {Guo}\ and\ \citenamefont
  {Oller}(2020)}]{Guo:2020pvt}%
  \BibitemOpen
  \bibfield  {author} {\bibinfo {author} {\bibfnamefont {Z.-H.}\ \bibnamefont
  {Guo}}\ and\ \bibinfo {author} {\bibfnamefont {J.}~\bibnamefont {Oller}},\
  }\href@noop {} {\  (\bibinfo {year} {2020})},\ \Eprint
  {http://arxiv.org/abs/2011.00978} {arXiv:2011.00978 [hep-ph]} \BibitemShut
  {NoStop}%
\bibitem [{\citenamefont {Zhu}\ \emph {et~al.}(2020)\citenamefont {Zhu},
  \citenamefont {Guo}, \citenamefont {Zhang}, \citenamefont {Ma},\ and\
  \citenamefont {Li}}]{Zhu:2020snb}%
  \BibitemOpen
  \bibfield  {author} {\bibinfo {author} {\bibfnamefont {J.-W.}\ \bibnamefont
  {Zhu}}, \bibinfo {author} {\bibfnamefont {X.-D.}\ \bibnamefont {Guo}},
  \bibinfo {author} {\bibfnamefont {R.-Y.}\ \bibnamefont {Zhang}}, \bibinfo
  {author} {\bibfnamefont {W.-G.}\ \bibnamefont {Ma}}, \ and\ \bibinfo {author}
  {\bibfnamefont {X.-Q.}\ \bibnamefont {Li}},\ }\href@noop {} {\  (\bibinfo
  {year} {2020})},\ \Eprint {http://arxiv.org/abs/2011.07799} {arXiv:2011.07799
  [hep-ph]} \BibitemShut {NoStop}%
\bibitem [{\citenamefont {Cao}\ \emph {et~al.}(2020)\citenamefont {Cao},
  \citenamefont {Chen}, \citenamefont {Qi},\ and\ \citenamefont
  {Zheng}}]{Cao:2020gul}%
  \BibitemOpen
  \bibfield  {author} {\bibinfo {author} {\bibfnamefont {Q.-F.}\ \bibnamefont
  {Cao}}, \bibinfo {author} {\bibfnamefont {H.}~\bibnamefont {Chen}}, \bibinfo
  {author} {\bibfnamefont {H.-R.}\ \bibnamefont {Qi}}, \ and\ \bibinfo {author}
  {\bibfnamefont {H.-Q.}\ \bibnamefont {Zheng}},\ }\href@noop {} {\  (\bibinfo
  {year} {2020})},\ \Eprint {http://arxiv.org/abs/2011.04347} {arXiv:2011.04347
  [hep-ph]} \BibitemShut {NoStop}%
\bibitem [{\citenamefont {liu}\ \emph {et~al.}(2020)\citenamefont {liu},
  \citenamefont {Liu}, \citenamefont {Zhong},\ and\ \citenamefont
  {Zhao}}]{liu:2020eha}%
  \BibitemOpen
  \bibfield  {author} {\bibinfo {author} {\bibfnamefont {M.-S.}\ \bibnamefont
  {liu}}, \bibinfo {author} {\bibfnamefont {F.-X.}\ \bibnamefont {Liu}},
  \bibinfo {author} {\bibfnamefont {X.-H.}\ \bibnamefont {Zhong}}, \ and\
  \bibinfo {author} {\bibfnamefont {Q.}~\bibnamefont {Zhao}},\ }\href@noop {}
  {\  (\bibinfo {year} {2020})},\ \Eprint {http://arxiv.org/abs/2006.11952}
  {arXiv:2006.11952 [hep-ph]} \BibitemShut {NoStop}%
\bibitem [{\citenamefont {Gong}\ \emph {et~al.}(2020)\citenamefont {Gong},
  \citenamefont {Du}, \citenamefont {Zhou}, \citenamefont {Zhao},\ and\
  \citenamefont {Zhong}}]{Gong:2020bmg}%
  \BibitemOpen
  \bibfield  {author} {\bibinfo {author} {\bibfnamefont {C.}~\bibnamefont
  {Gong}}, \bibinfo {author} {\bibfnamefont {M.-C.}\ \bibnamefont {Du}},
  \bibinfo {author} {\bibfnamefont {B.}~\bibnamefont {Zhou}}, \bibinfo {author}
  {\bibfnamefont {Q.}~\bibnamefont {Zhao}}, \ and\ \bibinfo {author}
  {\bibfnamefont {X.-H.}\ \bibnamefont {Zhong}},\ }\href@noop {} {\  (\bibinfo
  {year} {2020})},\ \Eprint {http://arxiv.org/abs/2011.11374} {arXiv:2011.11374
  [hep-ph]} \BibitemShut {NoStop}%
\bibitem [{\citenamefont {Liu}\ \emph {et~al.}(2020)\citenamefont {Liu},
  \citenamefont {Xie},\ and\ \citenamefont {Geng}}]{Liu:2020nil}%
  \BibitemOpen
  \bibfield  {author} {\bibinfo {author} {\bibfnamefont {M.-Z.}\ \bibnamefont
  {Liu}}, \bibinfo {author} {\bibfnamefont {J.-J.}\ \bibnamefont {Xie}}, \ and\
  \bibinfo {author} {\bibfnamefont {L.-S.}\ \bibnamefont {Geng}},\ }\href
  {\doibase 10.1103/PhysRevD.102.091502} {\bibfield  {journal} {\bibinfo
  {journal} {Phys. Rev. D}\ }\textbf {\bibinfo {volume} {102}},\ \bibinfo
  {pages} {091502} (\bibinfo {year} {2020})},\ \Eprint
  {http://arxiv.org/abs/2008.07389} {arXiv:2008.07389 [hep-ph]} \BibitemShut
  {NoStop}%
\bibitem [{\citenamefont {Valderrama}(2012)}]{Valderrama:2012jv}%
  \BibitemOpen
  \bibfield  {author} {\bibinfo {author} {\bibfnamefont {M.}~\bibnamefont
  {Valderrama}},\ }\href {\doibase 10.1103/PhysRevD.85.114037} {\bibfield
  {journal} {\bibinfo  {journal} {Phys. Rev. D}\ }\textbf {\bibinfo {volume}
  {85}},\ \bibinfo {pages} {114037} (\bibinfo {year} {2012})},\ \Eprint
  {http://arxiv.org/abs/1204.2400} {arXiv:1204.2400 [hep-ph]} \BibitemShut
  {NoStop}%
\bibitem [{\citenamefont {Lu}\ \emph {et~al.}(2019)\citenamefont {Lu},
  \citenamefont {Geng},\ and\ \citenamefont {Valderrama}}]{Lu:2017dvm}%
  \BibitemOpen
  \bibfield  {author} {\bibinfo {author} {\bibfnamefont {J.-X.}\ \bibnamefont
  {Lu}}, \bibinfo {author} {\bibfnamefont {L.-S.}\ \bibnamefont {Geng}}, \ and\
  \bibinfo {author} {\bibfnamefont {M.~P.}\ \bibnamefont {Valderrama}},\ }\href
  {\doibase 10.1103/PhysRevD.99.074026} {\bibfield  {journal} {\bibinfo
  {journal} {Phys. Rev.}\ }\textbf {\bibinfo {volume} {D99}},\ \bibinfo {pages}
  {074026} (\bibinfo {year} {2019})},\ \Eprint
  {http://arxiv.org/abs/1706.02588} {arXiv:1706.02588 [hep-ph]} \BibitemShut
  {NoStop}%
\bibitem [{\citenamefont {Lewis}\ and\ \citenamefont
  {Woloshyn}(2009)}]{Lewis:2008fu}%
  \BibitemOpen
  \bibfield  {author} {\bibinfo {author} {\bibfnamefont {R.}~\bibnamefont
  {Lewis}}\ and\ \bibinfo {author} {\bibfnamefont {R.~M.}\ \bibnamefont
  {Woloshyn}},\ }\href {\doibase 10.1103/PhysRevD.79.014502} {\bibfield
  {journal} {\bibinfo  {journal} {Phys. Rev.}\ }\textbf {\bibinfo {volume}
  {D79}},\ \bibinfo {pages} {014502} (\bibinfo {year} {2009})},\ \Eprint
  {http://arxiv.org/abs/0806.4783} {arXiv:0806.4783 [hep-lat]} \BibitemShut
  {NoStop}%
\bibitem [{\citenamefont {Cho}(1993)}]{Cho:1992cf}%
  \BibitemOpen
  \bibfield  {author} {\bibinfo {author} {\bibfnamefont {P.~L.}\ \bibnamefont
  {Cho}},\ }\href {\doibase 10.1016/0550-3213(94)90522-3,
  10.1016/0550-3213(93)90263-O} {\bibfield  {journal} {\bibinfo  {journal}
  {Nucl. Phys.}\ }\textbf {\bibinfo {volume} {B396}},\ \bibinfo {pages} {183}
  (\bibinfo {year} {1993})},\ \bibinfo {note} {[Erratum: Nucl.
  Phys.B421,683(1994)]},\ \Eprint {http://arxiv.org/abs/hep-ph/9208244}
  {arXiv:hep-ph/9208244 [hep-ph]} \BibitemShut {NoStop}%
\bibitem [{\citenamefont {Baru}\ \emph {et~al.}(2016)\citenamefont {Baru},
  \citenamefont {Epelbaum}, \citenamefont {Filin}, \citenamefont {Hanhart},
  \citenamefont {Mei\ss{}ner},\ and\ \citenamefont {Nefediev}}]{Baru:2016iwj}%
  \BibitemOpen
  \bibfield  {author} {\bibinfo {author} {\bibfnamefont {V.}~\bibnamefont
  {Baru}}, \bibinfo {author} {\bibfnamefont {E.}~\bibnamefont {Epelbaum}},
  \bibinfo {author} {\bibfnamefont {A.}~\bibnamefont {Filin}}, \bibinfo
  {author} {\bibfnamefont {C.}~\bibnamefont {Hanhart}}, \bibinfo {author}
  {\bibfnamefont {U.-G.}\ \bibnamefont {Mei\ss{}ner}}, \ and\ \bibinfo {author}
  {\bibfnamefont {A.}~\bibnamefont {Nefediev}},\ }\href {\doibase
  10.1016/j.physletb.2016.10.008} {\bibfield  {journal} {\bibinfo  {journal}
  {Phys. Lett. B}\ }\textbf {\bibinfo {volume} {763}},\ \bibinfo {pages} {20}
  (\bibinfo {year} {2016})},\ \Eprint {http://arxiv.org/abs/1605.09649}
  {arXiv:1605.09649 [hep-ph]} \BibitemShut {NoStop}%
\bibitem [{\citenamefont {Prelovsek}\ \emph {et~al.}(2020)\citenamefont
  {Prelovsek}, \citenamefont {Collins}, \citenamefont {Mohler}, \citenamefont
  {Padmanath},\ and\ \citenamefont {Piemonte}}]{Prelovsek:2020eiw}%
  \BibitemOpen
  \bibfield  {author} {\bibinfo {author} {\bibfnamefont {S.}~\bibnamefont
  {Prelovsek}}, \bibinfo {author} {\bibfnamefont {S.}~\bibnamefont {Collins}},
  \bibinfo {author} {\bibfnamefont {D.}~\bibnamefont {Mohler}}, \bibinfo
  {author} {\bibfnamefont {M.}~\bibnamefont {Padmanath}}, \ and\ \bibinfo
  {author} {\bibfnamefont {S.}~\bibnamefont {Piemonte}},\ }\href@noop {} {\
  (\bibinfo {year} {2020})},\ \Eprint {http://arxiv.org/abs/2011.02542}
  {arXiv:2011.02542 [hep-lat]} \BibitemShut {NoStop}%
\bibitem [{\citenamefont {Wang}\ \emph
  {et~al.}(2020{\natexlab{c}})\citenamefont {Wang}, \citenamefont {Li},
  \citenamefont {Liang},\ and\ \citenamefont {Oset}}]{Wang:2020elp}%
  \BibitemOpen
  \bibfield  {author} {\bibinfo {author} {\bibfnamefont {E.}~\bibnamefont
  {Wang}}, \bibinfo {author} {\bibfnamefont {H.-S.}\ \bibnamefont {Li}},
  \bibinfo {author} {\bibfnamefont {W.-H.}\ \bibnamefont {Liang}}, \ and\
  \bibinfo {author} {\bibfnamefont {E.}~\bibnamefont {Oset}},\ }\href@noop {}
  {\  (\bibinfo {year} {2020}{\natexlab{c}})},\ \Eprint
  {http://arxiv.org/abs/2010.15431} {arXiv:2010.15431 [hep-ph]} \BibitemShut
  {NoStop}%
\bibitem [{\citenamefont {Ding}\ \emph {et~al.}(2020)\citenamefont {Ding},
  \citenamefont {Jiang},\ and\ \citenamefont {He}}]{Ding:2020dio}%
  \BibitemOpen
  \bibfield  {author} {\bibinfo {author} {\bibfnamefont {Z.-M.}\ \bibnamefont
  {Ding}}, \bibinfo {author} {\bibfnamefont {H.-Y.}\ \bibnamefont {Jiang}}, \
  and\ \bibinfo {author} {\bibfnamefont {J.}~\bibnamefont {He}},\ }\href@noop
  {} {\  (\bibinfo {year} {2020})},\ \Eprint {http://arxiv.org/abs/2011.04980}
  {arXiv:2011.04980 [hep-ph]} \BibitemShut {NoStop}%
\bibitem [{\citenamefont {Dai}\ \emph {et~al.}(2020)\citenamefont {Dai},
  \citenamefont {Toledo},\ and\ \citenamefont {Oset}}]{Dai:2020yfu}%
  \BibitemOpen
  \bibfield  {author} {\bibinfo {author} {\bibfnamefont {L.}~\bibnamefont
  {Dai}}, \bibinfo {author} {\bibfnamefont {G.}~\bibnamefont {Toledo}}, \ and\
  \bibinfo {author} {\bibfnamefont {E.}~\bibnamefont {Oset}},\ }\href {\doibase
  10.1140/epjc/s10052-020-8058-8} {\bibfield  {journal} {\bibinfo  {journal}
  {Eur. Phys. J. C}\ }\textbf {\bibinfo {volume} {80}},\ \bibinfo {pages} {510}
  (\bibinfo {year} {2020})},\ \Eprint {http://arxiv.org/abs/2004.05204}
  {arXiv:2004.05204 [hep-ph]} \BibitemShut {NoStop}%
\bibitem [{\citenamefont {Tornqvist}(1994)}]{Tornqvist:1993ng}%
  \BibitemOpen
  \bibfield  {author} {\bibinfo {author} {\bibfnamefont {N.~A.}\ \bibnamefont
  {Tornqvist}},\ }\href {\doibase 10.1007/BF01413192} {\bibfield  {journal}
  {\bibinfo  {journal} {Z. Phys.}\ }\textbf {\bibinfo {volume} {C61}},\
  \bibinfo {pages} {525} (\bibinfo {year} {1994})},\ \Eprint
  {http://arxiv.org/abs/hep-ph/9310247} {arXiv:hep-ph/9310247 [hep-ph]}
  \BibitemShut {NoStop}%
\bibitem [{\citenamefont {Liu}\ \emph {et~al.}(2018)\citenamefont {Liu},
  \citenamefont {Wu}, \citenamefont {Xie}, \citenamefont {Pavon~Valderrama},\
  and\ \citenamefont {Geng}}]{Liu:2018bkx}%
  \BibitemOpen
  \bibfield  {author} {\bibinfo {author} {\bibfnamefont {M.-Z.}\ \bibnamefont
  {Liu}}, \bibinfo {author} {\bibfnamefont {T.-W.}\ \bibnamefont {Wu}},
  \bibinfo {author} {\bibfnamefont {J.-J.}\ \bibnamefont {Xie}}, \bibinfo
  {author} {\bibfnamefont {M.}~\bibnamefont {Pavon~Valderrama}}, \ and\
  \bibinfo {author} {\bibfnamefont {L.-S.}\ \bibnamefont {Geng}},\ }\href
  {\doibase 10.1103/PhysRevD.98.014014} {\bibfield  {journal} {\bibinfo
  {journal} {Phys. Rev.}\ }\textbf {\bibinfo {volume} {D98}},\ \bibinfo {pages}
  {014014} (\bibinfo {year} {2018})},\ \Eprint
  {http://arxiv.org/abs/1805.08384} {arXiv:1805.08384 [hep-ph]} \BibitemShut
  {NoStop}%
\bibitem [{\citenamefont {Meng}\ \emph {et~al.}(2017)\citenamefont {Meng},
  \citenamefont {Li},\ and\ \citenamefont {Zhu}}]{Meng:2017fwb}%
  \BibitemOpen
  \bibfield  {author} {\bibinfo {author} {\bibfnamefont {L.}~\bibnamefont
  {Meng}}, \bibinfo {author} {\bibfnamefont {N.}~\bibnamefont {Li}}, \ and\
  \bibinfo {author} {\bibfnamefont {S.-L.}\ \bibnamefont {Zhu}},\ }\href
  {\doibase 10.1103/PhysRevD.95.114019} {\bibfield  {journal} {\bibinfo
  {journal} {Phys. Rev.}\ }\textbf {\bibinfo {volume} {D95}},\ \bibinfo {pages}
  {114019} (\bibinfo {year} {2017})},\ \Eprint
  {http://arxiv.org/abs/1704.01009} {arXiv:1704.01009 [hep-ph]} \BibitemShut
  {NoStop}%
\bibitem [{\citenamefont {Yang}\ \emph
  {et~al.}(2020{\natexlab{b}})\citenamefont {Yang}, \citenamefont {Meng},\ and\
  \citenamefont {Zhu}}]{Yang:2019rgw}%
  \BibitemOpen
  \bibfield  {author} {\bibinfo {author} {\bibfnamefont {B.}~\bibnamefont
  {Yang}}, \bibinfo {author} {\bibfnamefont {L.}~\bibnamefont {Meng}}, \ and\
  \bibinfo {author} {\bibfnamefont {S.-L.}\ \bibnamefont {Zhu}},\ }\href
  {\doibase 10.1140/epja/s10050-020-00028-9} {\bibfield  {journal} {\bibinfo
  {journal} {Eur. Phys. J. A}\ }\textbf {\bibinfo {volume} {56}},\ \bibinfo
  {pages} {67} (\bibinfo {year} {2020}{\natexlab{b}})},\ \Eprint
  {http://arxiv.org/abs/1906.04956} {arXiv:1906.04956 [hep-ph]} \BibitemShut
  {NoStop}%
\end{thebibliography}%

\end{document}